\documentclass[prb,aps,amsfonts,amssymb,floatfix,showpacs,superscriptaddress]{revtex4}
\usepackage{graphicx}
\usepackage{dcolumn}% Align table columns on decimal point
\usepackage{bm}% bold math
\usepackage{color} 
\usepackage{epsfig}
\usepackage{epstopdf}
\begin{document} %****
\title{%nccozx9\_LSCOE\_2D\_short7.tex: 
Condensation Bottleneck Driven by a Hidden Van Hove Singularity as the  Origin of Pseudogap Physics in Cuprate High-$T_c$ Superconductors}
\author{R.S. Markiewicz, I.G. Buda, P. Mistark, and A. Bansil}
\affiliation{ Physics Department, Northeastern University, Boston MA 02115, USA}
\begin{abstract}
We propose a new approach to understand the origin of the pseudogap in the cuprates.  The near-simultaneous softening of a large number of different $q$-bosons yields an ordering bottleneck, wherein the growth of magnetic correlations with decreasing temperature is anomalously slow, leading to extended ranges of short-range order.  This effect is not tied to the Fermi level, but driven by a Van Hove singularity (VHS) nesting that is strongest near a temperature $T_{VHS}$ that scales as the energy separation between the Fermi energy and the energy of the VHS peak. By identifying $T_{VHS}$ as the pseudogap onset temperature $T^*$, we explain many characteristic features of 
the pseudogap, including the observed transport anomalies and the termination of the pseudogap phase when $E_{VHS}$ crosses the Fermi level. The condensation bottleneck (CB) provides a new pathway for understanding strong correlation effects in the cuprates.  We find that LSCO lies close to an anomalous disorder-free {\it spin glass} quantum critical point, where the frustration is due to the CB.  To study this point we develop an approach to interpolate between different cuprates.

\end{abstract} 
\maketitle
%\pacs{PACS numbers~:~~74.20.Mn,74.72.-h, 71.45.Lr, 74.50.+r }

%****
%\narrowtext
%****
\section{Introduction}

Evidence is growing that the `pseudogap phase' found in cuprates is actually home to one or more competing phases, including a variety of stripe, spin-, or charge-density wave (S/CDW) phases.\textsuperscript{1-12}  The CDW phase, in particular, has stimulated considerable interest\textsuperscript{13-22}. Many of these phases, including superconductivity, seem to appear at temperatures well below the pseudogap temperature $T^*$, so the exact relation between the pseudogap and these other phases remains elusive.  
Indeed, the real puzzle is understanding why the pseudogap phase bears so little resemblance to a conventional phase transition.  Here we demonstrate that the pseudogap phenomenon arises from a {\it condensation bottleneck} (CB), where a large number of density waves (DWs) with similar $q$-vectors attempt to soften and condense at the same time, leading to anomalously low transition temperatures and extended ranges of short-range order, characteristic of the pseudogap phase.  We note that many treatments of correlation effects -- e.g., path-integral approaches -- begin by finding a mean-field order parameter at some particular $q$-vector -- a form of random-phase approximation (RPA) -- and then include the effect of fluctuations.  Then if several instabilities compete, the one with lowest free energy wins out.  In this case, the CB represents a particularly strong correlation effect, insofar as one cannot solve the problem one-$q$ at a time, but must account for the $q$-competition from the start, to account for the entropy associated with the $q$-manifold. We note that RPA-type approximations extend well beyond simple mean-field theories, arising in spin-fermion models and in Hertz-Millis\textsuperscript{23,24} type of quantum critical theories when the Stoner denominator is approximated by a $T$-independent Ornstein-Zernicke\textsuperscript{25} form, and in all cases they miss the essential and strongly $T$-dependent mode-coupling physics.  

Similar effects have arisen in the past.  In Overhauser's theory of CDWs in alkali metals,
\textsuperscript{26} the $q$-vectors for all points on a spherical Fermi surface (FS) become unstable simultaneously and cannot be handled one at a time.  These effects are often referred to as phonon entropy \textsuperscript{27}, or more generally as boson entropy, and can be analyzed via vertex corrections which take proper account of mode coupling\textsuperscript{28,29}.   In excitonic theories, the various DW instabilities represent condensation of some electronic boson at a single $q$-vector, with triplet excitons corresponding to SDWs and singlet excitons to CDWs\textsuperscript{30}.  Summing only bubble and ladder diagrams reproduces the RPA results with BCS-like ratios of gap to critical temperature, $2\Delta/k_BT_c$.  Going beyond ladders (e.g., Bethe-Salpeter equation) incorporates mode coupling with enhanced $\Delta/T_c$ ratios  (compare Refs.~[31] and~[32]; for a review of excitonic insulators, see Ref.~[30]).  A particularly interesting analogy is provided by the Bose condensation of excitons.  The mean-field transition is found to correspond to the temperature at which excitons are formed.  However, since the excitons are localized in real space, they are greatly spread in $q$.  When fluctuations are included, the real transition lies at a much lower $T$, when all excitons condense into the lowest $q$ state.  Here we develop a similar theory for the cuprates via a self-consistent renormalization calculation of the vertex corrections.  We find that the CB not only drives pseudogap physics, but can potentially  lead to a novel disorder-free spin-glass phase.  Similar effects are likely to be present in many other families of correlated materials. 

We find that the phase diagram of La$_{2-x}$Sr$_x$CuO$_4$ (LSCO) is different from that of other cuprates.  To explore the transition between them, we simplify the first principles dispersions to equivalent `reference family' dispersions, depending only on the three nearest neighbor hopping parameters, $t$, $t'$, and $t''$ (Supplementary Material Section~I).  We find that all cuprates fall along a single cut in $t'/t-t''/t$-space, namely $t''=-t'/2$, with LSCO characterized by a smaller value of $t'$.

\section{Results}

\subsection{Bosonic VHS and Origin of the Susceptibility Plateau}
%\subsubsection{Anomalous VHS Nesting}

A proper understanding of the CB involves two different issues: firstly, since Fermi surface (FS) nesting typically singles out only a few discrete $q$-vectors, what causes many $q$--vectors to soften together?  Secondly, a formalism needs to be developed to handle the effects of bottleneck.  Both of these issues are addressed here.  Remarkably, in cuprates we find that the underlying source of strong mode coupling can be traced back to the Lindhard susceptibility $\chi_0$.

In a typical calculation of classical or quantum phase transitions\textsuperscript{23,24}, material parameters are introduced via $\chi_0$, which is used to define an interacting susceptibility $\chi$.  In our calculation, $\chi_0$ is calculated from density functional theory (DFT) bands corrected by a GW self-energy [Methods Section], and $\chi$ is the resulting RPA susceptibility.  The phase transition then corresponds to the vanishing of the denominator of $\chi$ at frequency $\omega =0$ for some momentum $q$.  To properly incorporate fluctuation effects, this `Stoner denominator' is then typically reduced to Ornstein-Zernicke (OZ)\textsuperscript{25} form, 
\begin{eqnarray}
\chi ({\bf q},\omega )\sim{1\over q^2+\xi^{-2}+i(\omega /\omega_c)^z},
\label{eq:3}
\end{eqnarray}
in terms of various deviations from the critical point (in $q$, $\omega$, and a `tuning parameter' which is proportional to $\xi^{-2}$, where $\xi$ is the correlation length).  Here $z$ is a dynamic exponent and $q$ is measured from the ${\bf Q}$ where the susceptibility has a peak. An important result of our calculation is that in cuprates, this OZ form must be treated carefully, since the coefficients of the deviation parameters can be strong functions of temperature and doping.   In particular, we find that the susceptibility inverse curvature (coefficient of $q^2$) can diverge due to a competition between conventional Fermi-surface nesting and Van Hove singularity (VHS) nesting.

Whereas many properties of a Landau Fermi liquid are determined by their values at the Fermi level, the susceptibility is an exception, having both bulk and Fermi surface contributions.  While the FS part contributes a ridge to $\chi_0$ that is a map of the FS at $q=2k_F$\textsuperscript{34} (where $k_F$ is the Fermi wave vector), in the cuprates there is also an important bulk contribution, which provides a smoothly varying background, peaking at $(\pi,\pi)$ and giving rise to the near-$(\pi,\pi)$-plateau in the susceptibility.  This peak can shift the balance of the FS nesting to $q$-vectors closer to the peaks, and in special cases can lead to commensurate nesting away from the FS nesting vector, at exactly $(\pi,\pi)$.  Moreover, as $T$ increases, coherent FS features are washed out, leaving behind only the commensurate bulk contribution.  This peak is a bosonic VHS (b-VHS), the finite-$q$ analog of the Van Hove excitons found in optical spectra\textsuperscript{35}, but present in the intraband susceptibility. 
However, it is a ``hidden'' b-VHS.  Despite the fact that it is pinned to zero energy {\it independent of doping or hopping parameters}, it is hidden in the sense that the effective density-of-states (DOS) exactly at the b-VHS peak almost always vanishes.  Some consequences of this are discussed in Supplementary Materials Section~II.

\begin{figure}
\leavevmode
\rotatebox{0}{\scalebox{0.34}{\includegraphics{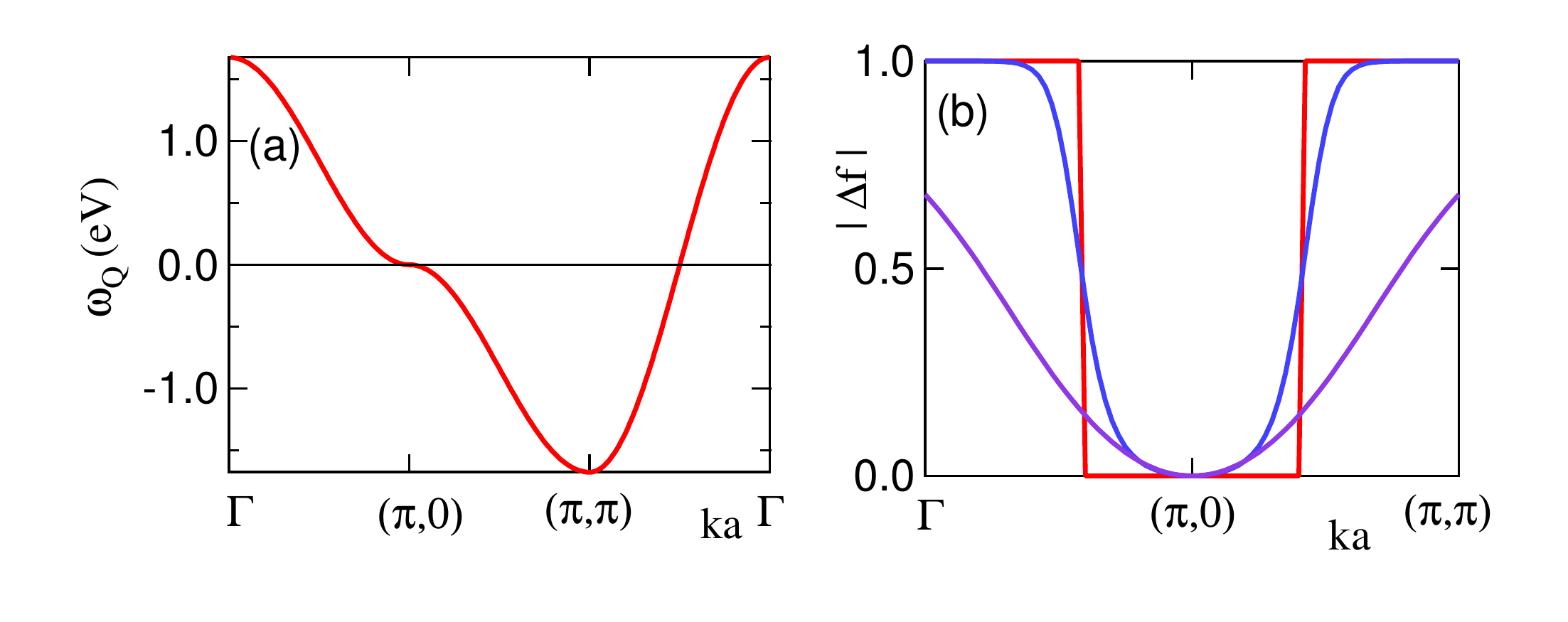}}}
\rotatebox{0}{\scalebox{0.54}{\includegraphics{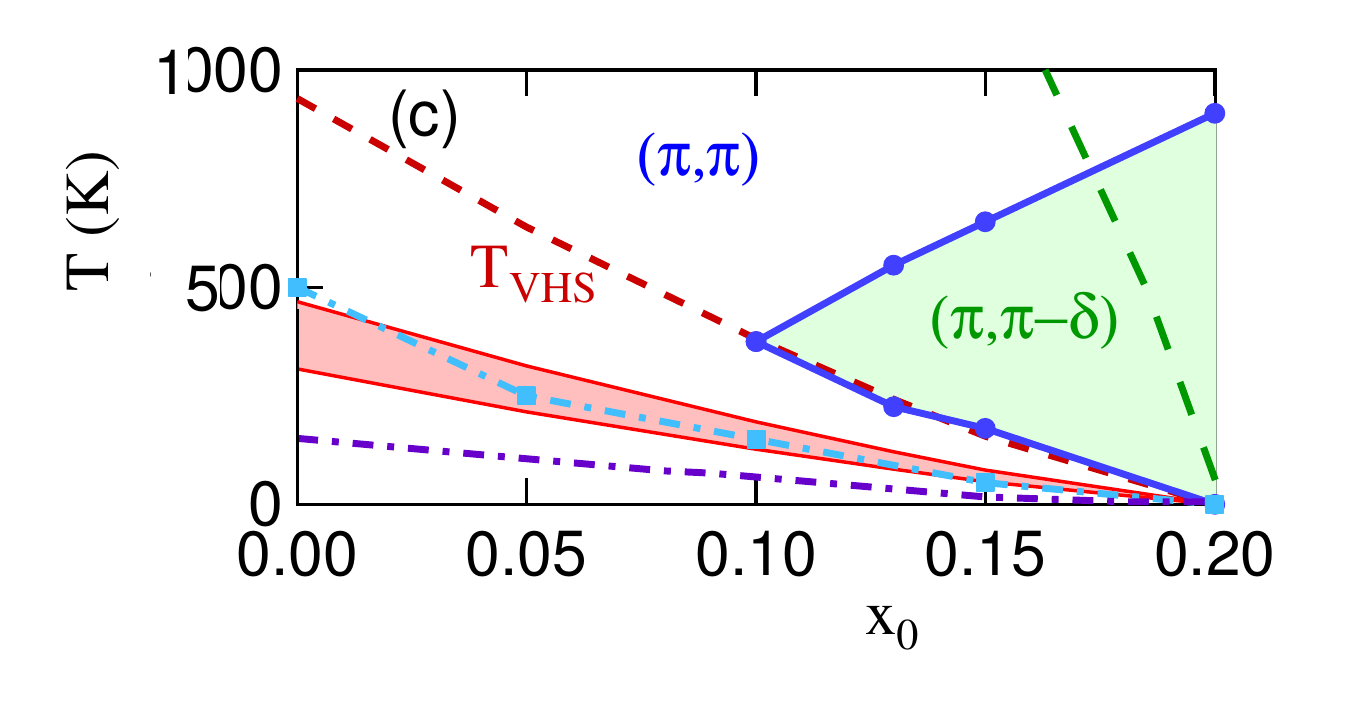}}}
%\rotatebox{0}{\scalebox{0.38}{\includegraphics{flsup1mg6a5_ttp_L8.eps}}}
\vskip0.5cm  
\caption{%(Color online) 
{\bf Coherent-incoherent crossover at the VHS.}
(a) Pair dispersion, $\omega_Q$, as a function of $k$, and (b) the corresponding weight $\Delta f$, at $T$ = 0K (red), 500K (blue), and 2000K (violet) for LSCO ( $x=0$).
(c)  Phase diagram for LSCO, showing VHS-related crossovers and regions where the susceptibility peaks at different $q$-vectors. Here $x_0$ is the doping at $T=0$ (plots are at constant $E_F$), but for qualitative purposes we can assume $x\simeq x_0$.  Dominant fluctuations are at commensurate $(\pi,\pi)$ (white shaded region) or incommensurate $(\pi,\pi-\delta)$ (green shaded region); crossovers are $T_{VHS}$ (red short-dashed line), DOS peak (green long-dashed line), coherent-incoherent crossover (pink shaded region), $T_{\gamma}$ (violet dot-dashed curve), and the position of the $(\pi,\pi)$ peak vs doping (light blue dot-dashed line) -- the approximate electron-hole symmetry point.
}
\label{fig:2c}
\end{figure}

The imaginary part of the susceptibility can be thought of as the DOS of electronic bosons, electron-hole (e-h) pairs, which may become excitons when a Coulomb interaction is turned on.  If the renormalized dispersion of a single electron is $\epsilon_k$ with wave vector $k$, then an e-h pair at wave vector $q$ has a dispersion $\omega_q(k)=\epsilon_{k+q}-\epsilon_k=-2\epsilon_{q-}(k)$, where $\epsilon_{q\pm}(k)=(\epsilon_k\pm\epsilon_{k+q})/2$, and a Pauli blocking factor $\Delta f_{k,q}=f(\epsilon_{k+q})-f(\epsilon_k)$.  Then the corresponding pair DOS is $D_q(\omega)=\sum_k\Delta f_{k,q}\delta(\omega- \omega_q(k))=\chi''_0(q,\omega)$.  For LSCO, the dominant pairs are those at $q=Q\equiv (\pi,\pi)$, the AF nesting vector.   The associated dispersion $\omega_Q(\pi,\pi)$, plotted in Fig.~\ref{fig:2c}(a), resembles the electronic dispersion $\epsilon_k$, but with an important distinction: it depends only on $\epsilon_{Q-}(k)$, whereas all of the hopping terms that shift the electronic VHS away from half filling ($t'$, $t''$) are contained in $\epsilon_{Q+}(k)$, i.e. the b-VHS is pinned at $\omega=0$.  Since $\chi_0''$ is an odd function of $\omega$, $\chi''_0(Q,\omega =0)=0$.  However, excitonic instabilities depend on a Stoner criterion, and hence on
\begin{equation}
\chi'_0(q,\omega=0)=2\int_0^{\infty} \frac{d\omega'}{\pi} \frac{\chi''_0(q,\omega')}{\omega'}.
\label{eq:E2}
\end{equation}

The FS contribution thus arises from $\omega'\sim 0$, while bulk contributions arise from peaks in the bosonic DOS, such as the b-VHS peak in $\chi''_0$ near $(\pi,\pi)$.  However, while the b-VHS is pinned at $\omega=0$, Fig.~\ref{fig:2c}(a), its weight vanishes at $T=0$, due to the Pauli blocking factor, $\Delta f _{k,Q}=0$ near $k=(\pi,0)$ at $T=0$, Fig.~\ref{fig:2c}(b). Finite $T$ restores weight, optimally near 1000K, although $\Delta f$ always vanishes exactly at $(\pi,0)$. In the Supplementary Materials, Section II, we deconvolve the near-$(\pi,\pi)$ susceptibility to show that it is a superposition of bulk and Fermi surface features.  The bulk feature is dominated by the b-VHS, while the FS feature smears out with increasing $T$ in a coherent-incoherent crossover, pink shaded region in Fig.~\ref{fig:2c}(c).  

To demonstrate the close similarity to VHS effects, we also plot two characteristic features of the VHS.  The crossover scales with both $k_BT_{VHS}=E_F-E_{VHS}$, dark red dashed curve in  Fig.~\ref{fig:2c}(c), and with  $T_{\gamma}$ (violet dot-dashed line), the temperature at which the Sommerfeld constant $\gamma=dS/dT$, has a peak.  Since the Fermi function evolves smoothly with $T$, we wxpect sharp Fermi surface features to wash out as $T$ is increased, but why should this coherent-incoherent crossover move to lower $T$ as doping is increased?  Because a source of entropy, the b-VHS, is moving closer to $E_F$.  To demonstrate the connection with entropy, we look at the Sommerfeld constant $\gamma$.  We calculate $\gamma$ from the electronic dispersion, assumng a paramagnetic phase, to avoid compications arising from phase transitions.  
At $T=0$, $\gamma$ is proportional to the DOS, and hence diverges when the VHS crosses $E_F$.  At finite $T$ the peak in $\gamma$ represents the excess  entropy associated with mode coupling.  The other curves in Fig.~\ref{fig:2c}(c) will be considered in the Discussion Section below.  

\subsection{Strong Mode Coupling Leads to an Extended Range of Short-range Order}

 In conventional mode-coupling calculations\textsuperscript{33,36-38}, the OZ parameters are assumed to be $T$- and doping-independent, and the resulting physics becomes quite simple.  For 2D materials, the Mermin-Wagner (MW) theorem\textsuperscript{39} is satisfied, and the mean-field transition at $T_{mf}$ turns into a pseudogap onset at $T^*\sim T_{mf}$, with a crossover to long-range order when interlayer coupling is strong enough -- in short, not much changes from the mean-field results.  However, when we incorporate realistic susceptibilities we find that this OZ form fails to properly account for the strong mode-coupling effects, leading to dramatically different results.  In particular, the entropic effects are encoded in a strong $T$-dependence of the model parameters, which greatly slows down the growth of correlations.   This in turn leads to an extended $T$-range of short-range fluctuations, which is typical of pseudogap physics.  We find that, particularly at high $T$, the doping dependence reflects the evolution of the anomalous VH scattering.  We note that while the OZ form is used in quantum critical theory\textsuperscript{23,24}, it is explicitly stated that it is to be used only in a limited $T$-range, and only in the absence of FS nesting\textsuperscript{24}, both of which are violated here.

Mode coupling modifies the bare Coulomb interaction $U$, producing a vertex-renormalized effective $U$, $U_{sp}=\Gamma U$ with $\Gamma =1/(1+\lambda)$, while $\lambda$ is found self-consistently from [see Methods section below]
\begin{eqnarray}
\lambda =\Gamma TA_0\int_{U_c}^{\infty} dX_- { N_-(X_-)\over X_--U_{sp}}. 
\label{eq:5}
\end{eqnarray}
where $A_0 =12u/\chi_0({\bf Q_0},0)$, $\chi_0({\bf Q_0},0)=max_q[\chi_0'({\bf q},0)]$, $u$ is the mode-coupling parameter, and $U_c=1/\chi_0({\bf Q_0},0)$.  Here, $\chi_0^{-1}$, denoted by $X_-$, is the variable of integration and we introduce a corresponding susceptibility density of states (SDOS) $N_-$. In Section III of Supplementary Materials, we show that Eq.~\ref{eq:5} is closely related to excitonic Bose condensation, with $\lambda$ proportional to the effective number of bosons.

\begin{figure}
\leavevmode
\rotatebox{0}{\scalebox{0.44}{\includegraphics{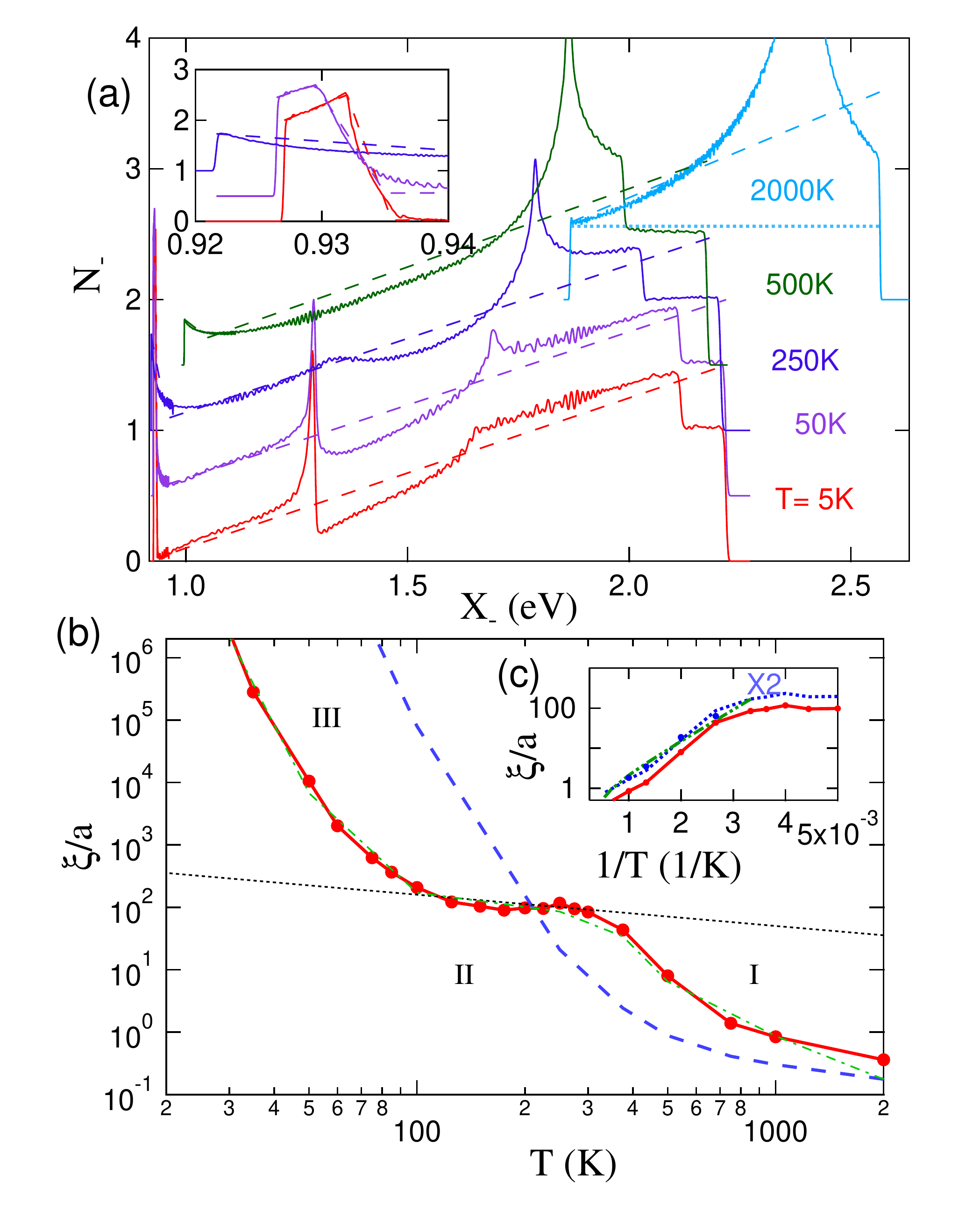}}}
%\rotatebox{0}{\scalebox{0.5}{\includegraphics{flsup1mg6a3_ttp2_G160}}}
%\rotatebox{0}{\scalebox{0.3}{\includegraphics{ModeC_3g}}}
%\rotatebox{0}{\scalebox{0.34}{\includegraphics{flsup1mg6a2w_exciton_L0.eps}}}
%\rotatebox{0}{\scalebox{0.38}{\includegraphics{The_figure_final_export.eps}}}
%\rotatebox{0}{\scalebox{0.34}{\includegraphics{Fig_e_h.eps}}}
\vskip0.5cm  
\caption{%(Color online) 
{\bf Structures in LSCO correlation length.}  (a) SDOS for undoped LSCO at several temperatures; light-blue dotted line gives OZ form of SDOS.  (b) Corresponding temperature dependence of correlation length $\xi_{th}$. For $x=0$, the red solid line with filled dots is $\xi_{th}$ calculated from the solid lines in (a), while the thin green dot-dashed line is based on the dashed lines in (a), showing that $\xi_{th}$ is relatively insensitive to the structures away from threshold.  In contrast, the blue dashed line is based on the shape of the SDOS at $T=2000K$, but shifted and renormalized to match the SDOS at lower $T$, illustrating sensitivity to the leading edge structure.  
%The violet dot-dashed curves correspond to $x=0.10$, with either $U=2~eV$ (thin line) or 1~eV (thick line).  
The black dotted line illustrates the scaling $\xi_{th}\propto T^{-1/2}$.
(c) Calculated $\xi_{th}$ replotted for $x=0$ (red solid line with filled circles) compared with $\xi$ (blue filled circles) and with experiment (green dot-dot-dashed line)\textsuperscript{40}; the blue dotted line is twice $\xi_{th}$.  
} 
\label{fig:3g}
\end{figure}

Figure~\ref{fig:3g} illustrates the profound effects that strong mode coupling has in LSCO, as well as the complete inability of the OZ approximation to capture this physics.  The SDOS, Fig.~\ref{fig:3g}(a), contains VHS-like features characteristic of conventional DOSs.  However, the singular behavior of Eq.~\ref{eq:5} involves only features near threshold, $U_c=min(X_-)$, which evolve strongly with $T$, see inset to Fig.~\ref{fig:3g}(a).  For $T>0$, the threshold behavior is always a step at $U_c$, indicative of a parabolic peak in $\chi_0$ with curvature inversely proportional to the step height. Thus for a qualitative understanding of the evolution of $\xi$ (Eq.~\ref{eq:3}) with $T$ we can assume an OZ form of $\chi$, but with a strongly $T$-dependent step height $A_2(T)$.  This allows a threshold correlation length $\xi_{th}$ (red solid lines in Figs.~\ref{fig:3g}(b) and~(c)) to be defined from Eq.~\ref{eq:3} as the inverse half-width in $q$, assuming that the curvature ($\propto A_2^{-1}$) is $q$-independent.  In reality, the curvature increases with $q$, and the correlation length $\xi$ obtained from the renormalized susceptibility half-width is typically a factor of 2 larger. Fig.~\ref{fig:3g}(c) compares the $x=0$ values of $\xi_{th}$ (solid red line) and $\xi$ (filled blue circles) vs $1/T$; the values of $\xi$ lie on the blue dotted curve, which represents twice $\xi_{th}$, and are in good agreement with experiment (green dot-dot-dashed line)\textsuperscript{40}.  

Figure~\ref{fig:3g}(b) illustrates how strong mode coupling slows down the correlation length divergence.  Undoped LSCO (red solid line) shows two regions I and III of exponential growth of $\xi_{th}$ with decreasing $T$, separated by an anomalous region II where $\xi_{th}$ actually decreases with decreasing $T$.  
%These effects are not restricted to 2D materials, and can contribute significantly to pseudogap physics in both 2D and 3D.  
We will not discuss the MW-like divergence at low $T$ (region III).  In the high-$T$ limit (region I), the leading-edge parabolic curvature is quite small, and if it were $T$-independent, as in the OZ approximation, the growth in $\xi_{th}$ would follow the blue dashed line (Eq.~\ref{eq:4} below), but thermal broadening causes the curvature to decrease with increasing $T$, leading to the faster growth of the red solid line.  The anomalous behavior in region II will be discussed next.

\subsection{Exploring parameter space}

To understand the origin of the CB in region II, it is necessary to explore hopping parameter space away from the physical cuprates.  To do this, we introduce the notion of reference states: states with simplified hopping parameters (only $t$, $t'$, and $t''$ nonzero) but which match the phase diagram of the real cuprates in a well-defined way [Supplementary Material Section I].  This allows us to tune the system between LSCO and Bi2212, and explore phase space beyond these limits.  We study two important cuts in $t'/t-t''/t$ space, a minimal cut ($t''=0$) and the Pavarini-Andersen [PA] cut ($t''=-t'/2$) -- the latter seems to best capture the physics of the cuprates.  By tuning $t'$ we unveil the origin of the CB as a localization-delocalization crossover tied to the crossover from $(\pi,\pi)$- to FS-nesting.  As a byproduct, we gain insight into why LSCO is so different from other cuprates, and how cuprates evolve from the pure Hubbard limit ($t'=0$).

\begin{figure}
\leavevmode
%\rotatebox{0}{\scalebox{0.44}{\includegraphics{flsup1mg6a3_ttp2_L27cd}}}
%\rotatebox{0}{\scalebox{0.38}{\includegraphics{The_figure_final_export.eps}}}
\rotatebox{0}{\scalebox{0.38}{\includegraphics{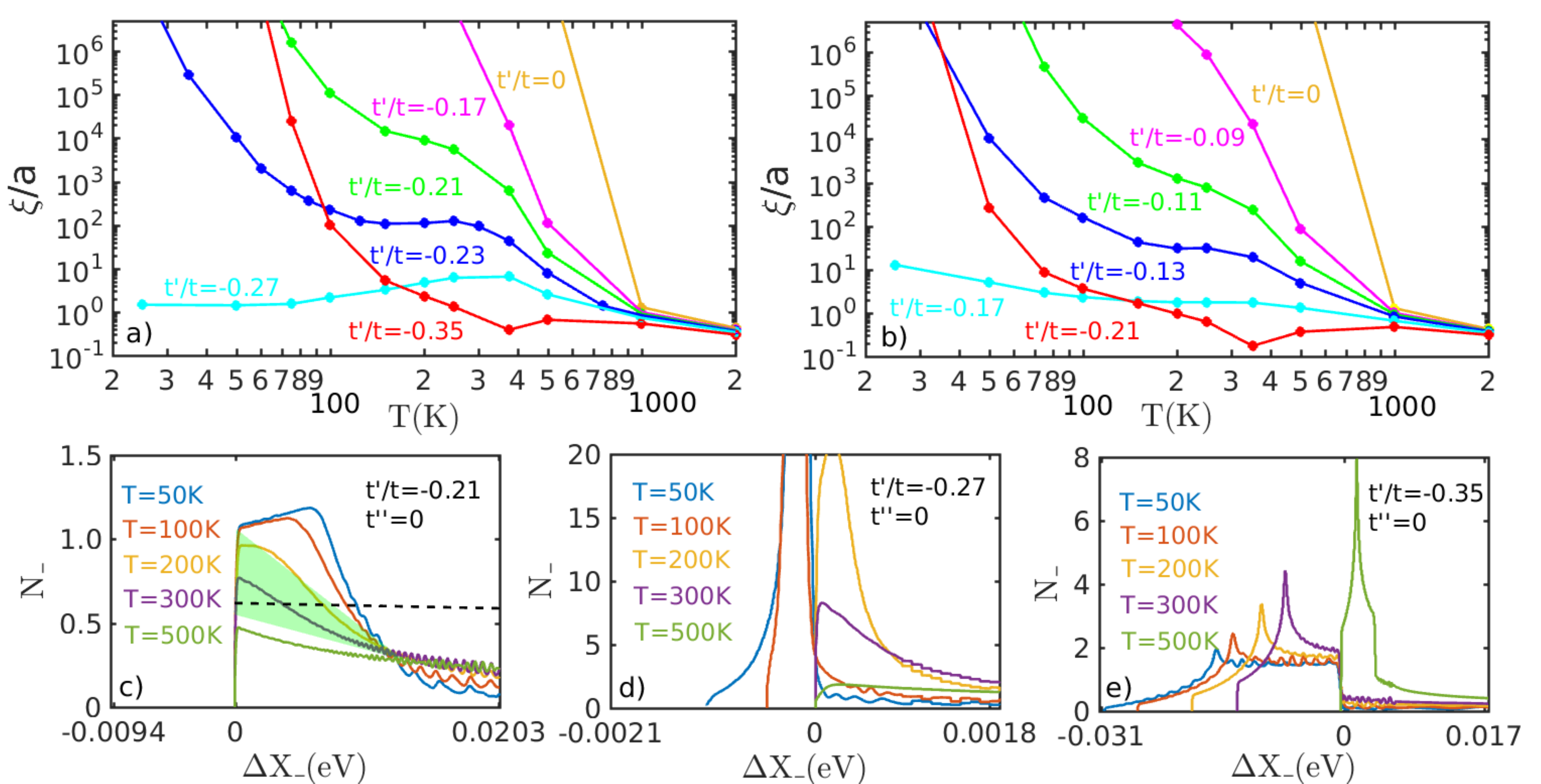}}}
\vskip0.5cm    
\caption{%(Color online) 
{\bf t' dependence of correlation bottleneck.}  (a)   $\xi_{th}$ for the minimal reference family at several values of $t'/t$ [see legend] and $t''=0$.  (b-d): SDOS at three values of $t'/t$ = -0.21 (b) [as in Fig.~\ref{fig:3g}(b)], -0.27 (c), and -0.35 (d).  All curves are shifted to line up the $(\pi,\pi)$ data at $\Delta X_- = X_- - X_-(\pi,\pi) =0$.  (e) $\xi_{th}$ for the PA reference family at several values of $t'/t$ [see legend] and $t''=-t'/2$.
} 
\label{fig:3g2}
\end{figure}
Figure~\ref{fig:3g2}(a) shows the $T$-evolution of $\xi_{th}$ for several values of $t'$ along the minimal cut, including the data of Fig.~\ref{fig:3g}(b).  For $t'/t >-0.17$, the system is characterized by commensurate (C) $(\pi,\pi)$ order with a finite Neel temperature $T_N\sim$1000K [the correlation length grows so rapidly that interlayer correlations will drive a transition to full 3D order].  Similarly, for $t'/t<-0.345$ there is incommensurate (I) $(\pi,\pi-\delta)$ order with $T_N$ about a factor of 10 smaller.  But for intermediate $t'/t$ the C-I transition is highly anomalous, with correlation length orders of magnitude smaller than expected.  Figure~\ref{fig:3g2}(b) shows that a similar evolution follows along the PA cut in parameter space.  The reason for this anomalous behavior can be seen by looking at the leading edge SDOS in the crossover regime, shown for three values of $t'$ along the minimal cut in Figs.~\ref{fig:3g2}(c)-(e).  For ease in viewing, these curves have been shifted to line up the SDOS at $(\pi,\pi)$ at all $T$.  It is seen that the anomalous collapse of $\xi_{th}$ is associated with a rapid growth of the step height $A_2$, culminating in a near-divergence at $t'_c=-0.27t$, where the leading edge curvature goes to zero.  This divergence coincides with the C-I crossover of the leading edge SDOS  [Supplementary Materials Section I].  Here many different $q$-vectors compete simultaneously, frustrating the divergence of any particular mode.  This is the electronic analog of McMillan's phonon entropy: if many phonons are simultaneously excited, the transition is suppressed to anomalously low temperatures.  Note that $\xi_{th}$ drops by 9 orders of magnitude at $T=200K$ when $t'/t$ changes from -0.17 to -0.27, then grows by a similar amount at 100K when $t'/t$ changes from -0.27 to -0.345. [Over this same range, the true $\xi$ will be frozen at the value corresponding to the half-width of the $(\pi,\pi)$ plateau.]  The green shaded region in Fig.~\ref{fig:3g2}(c) shows that the range of the anomalous growth II of $\xi$ in Fig.~\ref{fig:3g2}(a) coincides with the range of rapid growth of the SDOS leading edge; such behavior is absent if a $T$-independent OZ form (black dashed line) is assumed.  The divergence of the SDOS at $T=0$ signals that the system at $t'=-0.27t$ is highly anomalous, with an {\it infinitely degenerate ground state}.  This represents an anomalous form of spin glass arising in the absence of disorder, with the frustration arising from strong mode coupling.  
%It should be noted that when LSCO is doped, the commensurate AFM order rapidly disappears, being replaced by a low-$T$ spin glass phase.  
Note also that when the C-I transition is at $T>~300K$, $\xi(T)$ has a sharp downward cusp at the transition, while above the transition $\xi < a$ is strongly suppressed.

From the relationship between $\xi_{th}$ and the step height $A_2$, Eq.~\ref{eq:4} below, we see that a diverging $A_2$ will cause $\xi_{th}\rightarrow 0$.  The divergence arises when the susceptibility at $(\pi,\pi)$ crosses over from a maximum to a local minimum.  In the latter case, the maximum intensity of $\chi (q,0)$ is spread along a `ring' in $q$-space surrounding $(\pi,\pi)$, so that the inverse susceptibility resembles a `Mexican hat'.  Approximating the $\chi^{-1}$-dispersion by a Mexican hat form $\chi^{-1}_0-A_2q^2+A_4q^4$ leads to a threshold divergence $N_-\sim (\chi^{-1}-U_c)^{-1/2}$, Fig.~\ref{fig:3g}(f).  Note that the SDOS-divergence resembles a 1D VHS in the conventional DOS, even though here the 1-D direction is the radial direction away from $(\pi,\pi)$. The divergence of $\chi$ along a ring can be thought of as a 2D analog of Overhauser's effect\textsuperscript{26}.   Similar effects are found in Bi2201 [Supplementary Materials Section IV].  Finally, for $t'$ near but greater than $t'_c$, the divergence is avoided, but proximity to the C-I transition leads to the rapid growth of a narrow peak in the leading-edge SDOS, Fig.~\ref{fig:3g}(e).  This in turn can cause an approximate power-law growth of $\xi_{th}\propto T^{-1/2}$ (black dotted line in Fig.~\ref{fig:3g}(b)), as calculated in the Methods section below, although the sharp turn-on of structure in the SDOS actually leads to an anomalous decrease in $\xi_{th}$ with decreasing $T$.

%With hole doping, the plateau expands and region II is lost, leaving only the slow growth regime (thin dot-dashed violet line in Fig.~\ref{fig:3g}(b)).  Doping also leads to a smaller, screened $U$ (thick dot-dashed violet line in Fig.~\ref{fig:3g}(b))\textsuperscript{41,42}, further shifting the growth of $\xi_{th}$ to lower $T$.  

\section{Discussion}
\subsection{Cuprate Pseudogap}

Figure~\ref{fig:3g}(g) captures an essential aspect of pseudogap physics: an extended regime of phase space where correlations remain only short ranged.
The reason that the C-I transition is so anomalous is that it is also an {\it incoherent-coherent} transition, with the bare susceptibility for small $|t'|$ dominated, not by FS nesting, but by a broad peak at $(\pi,\pi)$ associated with a b-VHS, crossing over to conventional (coherent) FS nesting only for $t'<t'_c$ or for larger $x$.  Mode coupling associated with self-consistent vertex corrections leads to a regime where the local and extended behaviors become strongly entangled, leading to a collapse of the magnetic correlation length and very slow growth of magnetic fluctuations.  

While the CB was explained by tuning parameter space, the results bear a striking resemblance to the temperature and doping evolution of the pseudogap phase.  Experimentally, both LSCO and YBCO display C $(\pi,\pi)$ long-range AFM order at $x=0$, with $T_N$ dropping rapidly with doping below $x=0.03$, crossing over into a regime of I $(\pi,\pi-\delta)$ fluctuations and low-$T$ spin glass behavior.  Given the proximity of LSCO to the C-I transition, finding a quantitative model of its doping phase diagram may prove difficult.  We can however note one plausible scenario.  It has been predicted\textsuperscript{38} that a large Hubbard $U$ will renormalize $t'$ to smaller values at half filling, although the magnitude of the effect is debated.  If doping screens $U$, and restores a larger $|t'|$, it could drive LSCO across the anomalous regime II, causing $T_N$ to drop by an order of magnitude.   Moreover, Fig.~\ref{fig:3g}(g) shows that the C-I transition involves a highly disordered regime separating two well-ordered phases.  Since ordering tends to lower the free energy of the electronic system, the disordered regime may represent a state of high free energy, and doping across this region can lead to a regime of [nanoscale] phase separation (NPS), which in LSCO is manifest as the stripe phase.
Hence, a model of the C-I transition, incorporating effects of NPS and disorder, could provide a good description of LSCO.

This broad $T$- and doping-regime of only short-range SDW order is a key characteristic of pseudogap physics.   Indeed, $T_{VHS}$ is typically close to the measured pseudogap temperature $T^*$ in most cuprates.   Identifying $T^*$ with $T_{VHS}$ can explain a number of puzzling features.  In particular, it has been found that the pseudogaps in Bi2201 and Bi2212 terminate when $T_{VHS}\rightarrow 0$, i.e., at the conventional VHS,\textsuperscript{43-47} similar to the crossovers seen in Fig.~\ref{fig:2c}(c), as well as in Fig.~1 %~\ref{fig:009},
of the Supplementary Materials.
  
%Notably, a similar effect is found in pnictides[Ref].
%\subsubsection{Transport Anomalies}

The presence of an excitonic VHS contribution to the incoherent susceptibility has a number of further consequences for cuprate physics.  First, the VHS onset near $T_{VHS}$ combined with the C-I 
%incoherent-to-coherent 
crossover at $T_{coh}\sim T_{VHS}/3$ can explain the anomalous transport properties found near the pseudogap.  Thus, the pink shaded region in Fig.~\ref{fig:2c}(c) shows the crossover from incoherent susceptibility dominated by the VHS at high-$T$ to coherent, FS-dominated susceptibility at low-$T$.  This parallels transport, where for $T>T^*$, the resistivity $\rho$ varies linearly with $T$,\textsuperscript{9}  behavior expected near a VHS\textsuperscript{46}.  For lower $T$, the resistivity is mixed, but $\rho\sim T^2$, as expected for a coherent Fermi liquid, is found below a $T_{coh}<T_{VHS}$.  This picture bears a resemblance to the Barzykin-Pines model of the cuprate pseudogap,\textsuperscript{47} identifying $T_{VHS}$ and $T_{VHS}/3$ with $T^*$ and $T^*/3$ in their model.  The underlying physics of the incoherent-to-coherent crossover in their model is related to Kondo lattice physics\textsuperscript{48}, with the VHS peak standing in for the Kondo resonance (see Supplementary Materials Section~V).  This raises the question of whether a similar mode-coupling calculation in heavy-fermion compounds could lead to a similar anomalous entanglement at the f-electron C-I transition.

The excitonic instability should be maximal when the susceptibility is approximately electron-hole (e-h) symmetrical in doping, and at $T=0$ this happens when the VHS is at the Fermi level.\textsuperscript{34}  At finite $T$, the e-h symmetrical point continues to coincide with a peak in the $(\pi,\pi)$ susceptibility at doping $x_p$, but $x_p$ shows a remarkably rapid evolution with $T$ towards half-filling (light-blue dot-dashed line in Fig.~\ref{fig:2c}(c)), nearly coinciding with the coherent-incoherent crossover.  We interpret this temperature dependence as follows:  The parameters $t'$ and $t''$ are relevant perturbations shifting the VHS from the pure Hubbard model value at $x=0$ where $t'=t''=0$.  When $k_BT>|t'|$, these perturbations become irrelevant, so that the effective VHS becomes electron-hole symmetric at $x=0$.  Lastly, the lower branch of the $(\pi,\pi)-(\pi,\pi-\delta)$ commensurate-incommensurate transition (green shaded region in Fig.~\ref{fig:2c}(c)) also scales, following $T_{VHS}$ very closely in this simple ($t-t'$ only) model.  Indeed, at $T=0$, this transition corresponds to the Fermi energy falling off of the $(\pi,\pi)$ plateau, providing another indication that the effective VHS is shifting towards half-filling as $T$ increases. In Supplemental Material Section II we show that this is another consequence of Pauli blocking.

The rapid thermal evolution of the $(\pi,\pi)$-VHS peak should be contrasted with the much smaller change in conventional VHS effects found near $\Gamma$. In particular, the DOS peak is dominated by near-FS physics, and hence displays a much weaker doping dependence, green long-dashed line in Fig.~\ref{fig:2c}(c).

An additional consequence is that in cuprates with small $t'$ hopping, such as LSCO, the anomalous VHS susceptibility can dominate even at $T=0$, leading to strong deviations from FS-nesting.  Thus, in most of Fig.~\ref{fig:2c}(c), the peak susceptibility is associated with VHS nesting at $q=(\pi,\pi)$, and only in an intermediate $T$-regime is FS nesting is found at $(\pi,\pi-\delta)$ (green shaded region). The resulting VHS-FS nesting competition plays a strong role in underdoped LSCO, associated with retrograde correlation length change with $T$, region II of Fig.~\ref{fig:3g}(b), which is a signal of proximity to a novel disorder-free spin-glass QCP.  Notably, in LSCO commensurate $(\pi,\pi)$ order disappears rapidly by $\sim 2\%$ doping, being replaced by incommensurate magnetic fluctuations and low-$T$ spin-glass effects, while neutron scattering has found that doped LSCO is close to a magnetic QCP.\textsuperscript{49}  
 The very different situation in most other cuprates is discussed in Sections~I,~IV of Supplementary Materials.

\subsection{Strong coupling physics}

We recall that our self-energy formalism\textsuperscript{50} is able to reproduce most spectral features of the insulating cuprates in terms of a $(\pi,\pi)$ ordered phase.  We reproduce not only the photoemission dispersions, limited to the lower Hubbard bands, but optical and x-ray spectra which depend sensitively on the Mott gap.  
In a related 3-band model, we reproduced the Zhang-Rice result that the first doped holes are predominantly oxygen character,  and our overall dispersions at half filling agree with [subsequent] DMFT results at least as well as DMFT results from different groups agree. [See further Supplementary Materials VI.] The problem with our earlier calculations is that they predict long-range $(\pi,\pi)$ AFM order at too high $T$.  The new self-consistent renormalization calculations have only short range order, in which case the upper and lower Hubbard band dispersion is reproduced, with a broadening $\sim 1/\xi$.\textsuperscript{33}

An antiferromagnet is considered to be weakly coupled if the Neel temperature $T_N$ increases with $U$, and strongly coupled if $T_N$ decreases with increasing $U$ [$T_N\sim J=4 t^2/U$].    Since this is a finite $T$ criterion, it is sensitive to the boson entropy effects we have been discussing.  Indeed, for large $U$ the system is nearly localized and the physics is again reminiscent of Bose condensation.  The simultaneous softening of many magnetic modes can be readily demonstrated from a simple Hartree-Fock model of an antiferromagnet.   For arbitrary $q$, the resulting gap for large $U$ is
\begin{equation}
\Delta E = \sqrt{U^2+\epsilon_-^2 }\simeq U+J\tilde\epsilon_-^2,
\label{eq:0aa1}
\end{equation}
where $\epsilon_-=(\epsilon_k-\epsilon_{k+q})/2 =-2t\tilde\epsilon_-$.  Thus, the difference in energy between any two spin configurations is a quantity of order $J$, so that when $T\sim J$, the system gains entropy by mixing different $q$-states, and long-range AF order is destroyed.\textsuperscript{33}  While this effect can be recognized in HF, only a theory that properly accounts for the mode competition can resolve it.  The present mode coupling model properly accounts for this effect via the vertex renormalization factor $\lambda$, which is an integral over the RPA $\chi$.  This can be seen from Fig.~\ref{fig:3g}(c), where the measured $\xi$ in LSCO (green dot-dot-dashed line)\textsuperscript{40} is compared to our calculation (filled blue dots).  The experimental data are shown only above 300K, since at lower $T$ interlayer coupling drives a transition to long-range order.  In the Heisenberg model, $\xi$ is a function only of $T/J$, and these data have been used to measure the exchange $J$.  Hence, our mode coupling calculation successfully reproduces this strong coupling result, as was found earlier for electron-doped cuprates\textsuperscript{33}.  The above results suggest that the C-I, incoherent-coherent transition is also a localization-delocalization transition, with the mode coupling associated with small effective hopping.

For more insight into the strong coupling limit, we note that in the two-particle self-consistent approach\textsuperscript{51}, $\lambda$ is determined by a sum rule involving double occupancy, which is fixed by assuming 
$$U_{sp}/U=<n_{\uparrow}n_{\downarrow}>/(n/2)^2,$$
which leads to a saturation of $U_{sp}$ as $U\rightarrow\infty$ (or $<n_{\uparrow}n_{\downarrow}>\sim 1/U$).  In our calculation, this saturation arises naturally, since $U_{sp}$ can never exceed $U_c$.

Quantum critical points (QCPs) in strongly correlated materials are often discussed in terms of deconfined QCPs (DQCPs), involving a non-Landau transition between two types of competing order, where a new form of excitation emerges exactly at the DQCP.  This has been refined for cuprates into an underlying competition between Mott physics and Fermi liquid physics, masked by a low-energy order parameter of the FL\textsuperscript{52}.  This is an apt description of the current results, with the Mott physics evolving into a low-$T$ AFM and the FL to a spin-density wave (SDW), and a spin glass at the DQPT.  This confirms the finding from dynamical cluster approximation calculations of the strong role of the VHS in Mott physics.\textsuperscript{53}

\subsection{Conclusions}

In conclusion, we find that the pseudogap is driven by tendencies toward magnetic order at or near $q=(\pi,\pi)$.  Strong mode coupling effects drive the characteristic anomalies of the pseudogap, in particular the existence of a broad doping- and temperature-range of only short- to intermediate-range order.  Finally, the underlying cause of the strong mode coupling is localization – in particular the fact that the extended-to-localized transition is not a simple crossover but a competition between FS-dominated and non-FS dominated (VHS-dominated) physics.  This competition across a manifold of $q$-states leads to a condensation bottlneck, which requires a new formalism for dealing with multi-$q$ mode softening.  The close similarity to heavy Fermion physics should be noted.  

This can be restated slightly differently.  The original ($t$-only) Hubbard model is particularly difficult to solve, as three separate instabilities are simultaneously present at half filling: the Mott instability, FS nesting, and VHS nesting.  A finite $t'$ shifts the latter two to finite doping.  For large $|t'|$ the three instabilities become well separated, but for small $|t'|$ the two nesting instabilities overlap and compete, leading to frustration and pseudogap physics.  
Significantly, we approach the problem from the intermediate-coupling side, suggesting that the full crossover from half-filling to large doping could be explored.  

In turn, this sheds light on strong correlation effects in the cuprates, in particular Mott vs Slater physics.  The localization associated with Mott physics has several manifestations.  One is that coherent hopping is restricted to shorter range.  That is consistent with stronger effects found at smaller $t'$.  Particularly for Mott physics, a second aspect is the breakdown of hybridization. The Cu-O hybridization in the cuprates spoils Mott physics [half filling does not imply one electron on each Cu].  Zhang-Rice singlet formation can be considered as a breakdown of hybridization [pure oxygen states at the top of the lower magnetic band, pure Cu at the bottom of the upper magnetic band], and is found more generally for $(\pi,\pi)$ AFM fluctuations near half filling in a three-band model\textsuperscript{50}.  Hence the strong evolution with $t'$ from localized to extended physics, with only LSCO close to Mott physics.

Rice {\it et al.}\textsuperscript{54} recently noted: ``The need for theoretical methods to handle ... short range correlations ... is a key challenge for the future."  Along this line, the present approach appears to explain a number of anomalous features associated with the pseudogap, including a coherent crossover associated with a competition between two DW orders (see Supplementary Materials Section~IV).  In fact, we have incorporated an important ingredient for strong-coupling calculations: a model of short-range AF order from which a $t-J$ model can be derived.  

In Anderson's RVB picture, he envisaged a regime of Mott physics where the FS played a neglible role; instead, most experimental studies find clear evidence for a well-defined FS, with competing phases associated with FS nesting, and a QCP associated with FS reconstruction.  Our results have a strong bearing on these competing scenarios: in {\it parameter space}, $t'$ is a relevant parameter tuning the system away from quasilocalized physics near $t'=0$, where the FS has a negligible role, into a delocalized regime dominated by FS physics.  The crossover is marked by a regime of strong disorder, where the correlation length remains small down to very low $T$.  Most families of cuprates lie on the delocalized side of this crossover, and hence are most easily understood from a FL-type picture.  LSCO appears to be on the localized side of the crossover, but close to it, consistent with experiment.\textsuperscript{49}  We note that the disorder line may be hard to directly access: since ordering transitions tend to lower the free energy, the line will be a high-energy state with low energy states to either side of it, suggesting an instability to nanoscale phase separation.

The VHS has been predicted to play a significant role in many materials, particularly in lower dimensional systems where $\chi$ diverges, but clear evidence for this remains sparse.  Thus, VH nesting was introduced as a possible cause of CDWs in dichalcogenides\textsuperscript{55}, but this interpretation remains disputed. Even in VO$_2$, the striking metal-insulator transition has been found to be driven by large phonon entropy.\textsuperscript{56}   Clearly, one problem is that the VHS has been assumed to play a role only when it is near the Fermi level, whereas Fig.~\ref{fig:2c}(c) demonstrates that its influence extends over a much wider doping range.  
For those who think condensed matter physics lies on the surface of the Fermi sea, the VHS is the iceberg, lurking.

It will be interesting to examine how the present results are modified by 
%incorporation of GW-type self-consistency, and by 
effects of disorder and interlayer coupling. An important issue is the extent to which the total number of QPs is conserved. More specifically, what is the relation between the number of incoherent electrons $(1-Z)N$ and the effective number of excitons $N_{eff}$?  Proximity to a spin glass phase could explain the failure of recent attempts to create an `artificial cuprate’ from related nickelate compounds, and may suggest a path to improved analog states.

\section{Methods}

%\subsubsection{Mode Coupling Formalism}

Our calculation is a form of many-body perturbation theory (MBPT) based on Hedin's scheme.  The scheme involves four elements: electrons are described by Green's functions $G$ with DFT-based dispersions renormalized by a self-energy $\Sigma$; electronic bosons [electron-hole pairs] are described by a spectral weight [susceptibility] $\chi$ renormalized by vertex corrections $\Gamma$.  Neglecting vertex corrections, the self energy can be calculated as a convolution of $G$ and $W=U^2\chi$, the GW approximation.  This approach has been used to solve the energy gap problem in semiconductors, where the $\Gamma$ correction leads to excitons via the solution of a Bethe-Salpeter equation, and in extending DMFT calculations to incorporate more correlations [e.g., DMFT+GW, etc.].  Our approach here is to extend our previous GW calculations [quasiparticle-GW or QPGW\textsuperscript{50}] to include vertex corrections.  

In QPGW we introduce an auxiliary function $G_Z=Z/(\omega-\epsilon_{\bf k}^{QP})$, where the dressed, or QP dispersion is $\epsilon_{\bf k}^{QP}=Z\epsilon_{\bf k}^{DFT}$, and $\epsilon_{\bf k}^{DFT}$ is the bare, or DFT dispersion. $G_Z$ behaves like the Green's function of a Landau-type QP -- a free electron with renormalized parameters that describes the low-energy dressed electronic excitations.  However, this is a non-Fermi liquid type QP, since the frequency-integral of $Im(G_Z)$ is $Z$ and not unity.  That is, the $Z$-QP describes only the coherent part of the electronic dispersion, and is not in a 1:1 correspondence with the original electrons.  The importance of such a correction can be readily demonstrated.  Since a Z-QP has only the weight $Z$ of a regular electron, the susceptibility [a convolution of two $G$s] is weaker by a factor of $Z^2$ than an ordinary bare susceptibility.  To match this effect in the Stoner criterion requires introducing an effective $U_{eff}=ZU$.  In contrast, MBPT calculations in semiconductors typically set the GW-corrected Green's function to $G_{GW}^{-1}=\omega-\epsilon_{\bf k}^{QP}$, where $\epsilon_{\bf k}^{QP}$ is the average GW-renormalized dispersion\textsuperscript{57}, thereby missing the reduced spectral weight of the low-energy, coherent part of the band.

%When a new material is studied, models are made based on the experimental dispersion.  This leads to a susceptibility $\chi_{exp}$ of the same form as $\chi_Z$, but with much greater intensity.  , using the `experimental’ dispersion requires adding further empirical parameters.  For example,
%We hypothesize that in the low-energy sector the remaining spectral weight 1-Z is associated with quasi bosons describing dressed electron-hole pairs.  To test this hypothesis, we need to go beyond the GW limit of MBPT and study the effects of the vertex correction $\Gamma$.

We work in a purely magnetic sector, with a single Hubbard $U$ controlling all fluctuations; there is a competition in this model between near-nodal (NNN) and antinodal nesting (ANN) which mimics the SDW-CDW competition in cuprates, sharing the same nesting vectors\textsuperscript{22,34}.

%\section{Mode Coupling Formalism for Nonanalytic Susceptibilities}

The self-consistent parameter $\lambda$ is found from a Matsubara sum of the susceptibility
\begin{equation}
\lambda ={A_0T\over N}\sum_{{\bf q},i\omega_n}\chi ({\bf q},i\omega_n),
\label{eq:1}
\end{equation}
%where $V$ is the volume, 
where $N$ is the number of $q$-points, 
\begin{equation}
\chi ({\bf q},i\omega_n)={\chi_0({\bf q},i\omega_n)\over 1+\lambda 
-U\chi_0({\bf q},i\omega_n)},
\label{eq:1b} 
\end{equation}
and the summation in Eq.~\ref{eq:1} can be transformed:
\begin{eqnarray}
{T\over N}\sum_{{\bf q},i\omega_n}\chi ({\bf q},i\omega_n)&=&\int 
{d^2qa^2\over 4\pi^2} \int_{0}^{\infty}{d\omega\over\pi}coth({\omega\over 
2T})\chi''({\bf q},\omega +i\delta )
\nonumber \\
&\simeq&\hat\lambda_1+\hat\lambda_2T,
\label{eq:2}
\end{eqnarray}
with $a$ the in-plane lattice constant,
\begin{equation}
\hat\lambda_1=\int {d^2qa^2\over 4\pi^2}
\int_{0}^{\infty}{d\omega\over\pi}\chi''({\bf q},\omega),
\label{eq:2b}
\end{equation}
\begin{equation}
\hat\lambda_2=\int {d^2qa^2\over 4\pi^2}\chi' ({\bf q},0 ).
\label{eq:2c}
\end{equation}
The term 
$\hat\lambda_1$ introduces a small, nonsingular correction\textsuperscript{33} to 
Eq.~\ref{eq:1}, which we neglect.  

\begin{figure}
\leavevmode
%\rotatebox{0}{\scalebox{0.38}{\includegraphics[scale=1.3,keepaspectratio=true]{flsup1mg6a2_wt3_L0c.eps}}}
\rotatebox{0}{\scalebox{0.38}{\includegraphics{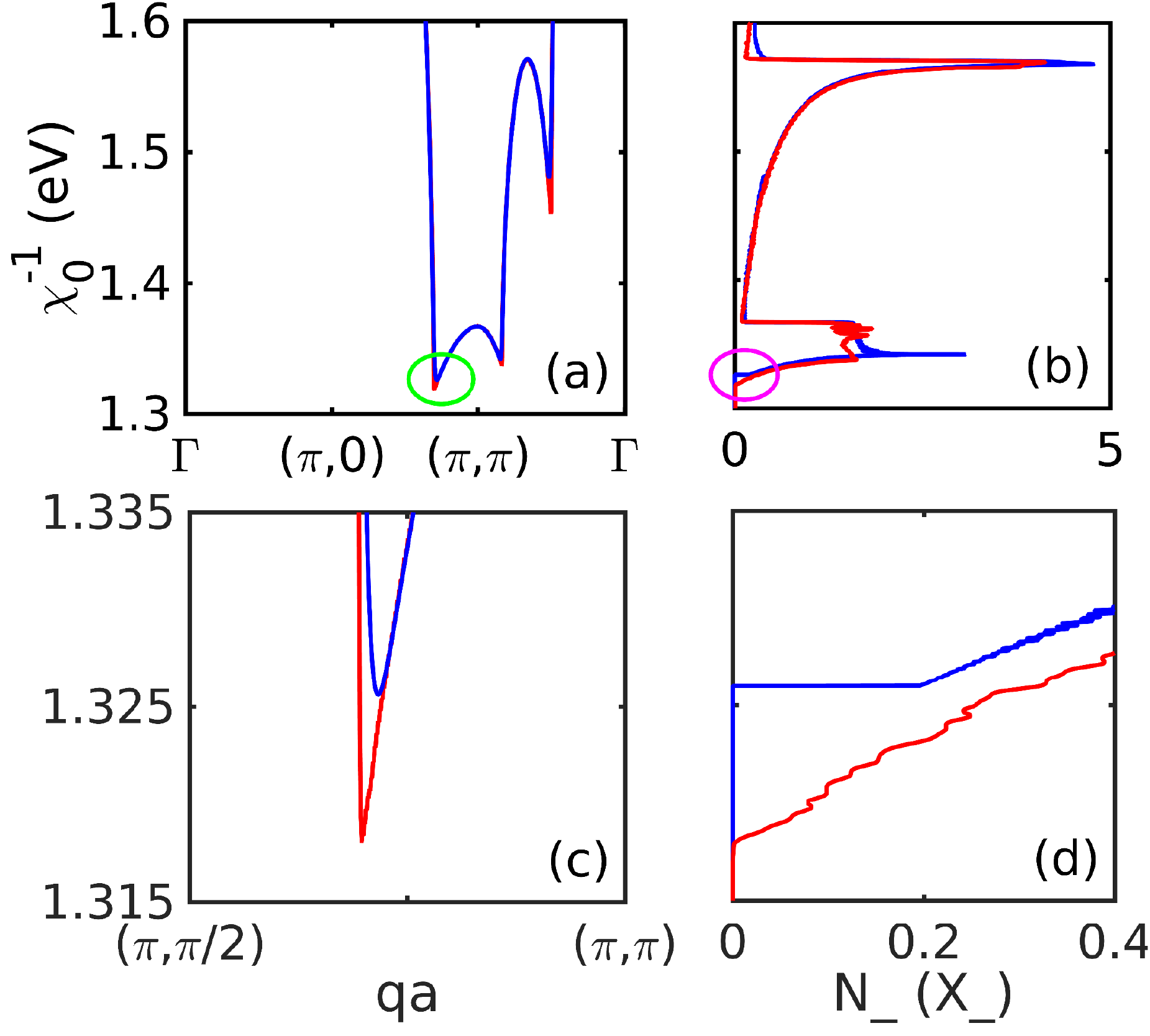}}}
\vskip0.5cm  
\caption{%(Color online) 
{\bf Origin of SDOS features.}
(a) Inverse susceptibility $\chi_0^{-1}({\bf q})$ for Bi2201  at doping
$x=0$.  Note that $U_c=min(\chi_0 '^{-1})$ (circled region). Curves are at $T$ = 10 (red) and 100K (blue).  (b) Corresponding susceptibility density of states (SDOS) $N_-(X_-)$, plotted horizontally, corresponding to the susceptibility of (a).  
(c,d) Blowups of circled regions in (a,b).
}
\label{fig:2b}
\end{figure}

In order to explore the role of susceptibility plateaus for realistic band dispersions, the self-consistency equation is evaluated numerically. 
For this purpose, we note that since $\chi_0^{-1}$ has dimensions of energy, a plot of $\chi_0^{-1}(q,0)$ resembles a dispersion map.  Hence we can define a SDOS: 
\begin{eqnarray}
\int {d^2qa^2\over 4\pi^2}=\int N_-(X_-) dX_-.
\label{eq:6}
\end{eqnarray}
Figure~\ref{fig:2b} illustrates how integrating over the inverse susceptibility, Fig.~\ref{fig:2b}(a), leads to the SDOS, Fig.~\ref{fig:2b}(b), at $x=0$ where the NNN plateau is dominant.  
Here we use hopping parameters appropriate to the DFT dispersion of Bi$_2$Sr$_2$CuO$_{6+x}$ (Bi2201), $t=419.5$, $t'=-108.2$ and $t''=54.1$~eV (Supplementary Materials Section~IV).
%, with a renormalization $Z=0.5$ included to simulate the experimental dispersion. 
This yields a doping phase diagram  that is qualitatively similar to that of most cuprates, except LSCO.  By comparing Figs~\ref{fig:2b}(a) and~\ref{fig:2b}(b), one can see how features in $\chi_0^{-1}$ translate into features in $N_-$.  Thus, the intense, flat-topped peak in $N_-$ at small values of $\chi_0^{-1}$ represents the NNN plateau.  Its broad leading edge (smaller $\chi_0^{-1}$) is controlled by anisotropy of the plateau edge between $(\pi,\pi-\delta)$ and $(\pi-\delta,\pi-\delta)$, while the sharp trailing edge corresponds to the local maximum of $\chi_0^{-1}$ at $(\pi,\pi)$.  For the ANN peak, its leading edge scarcely leaves any feature in $N_-$, but a local maximum translates into a large peak in $N_-$.  Finite temperature, $T=100K$ (blue line) rounds off the cusp in $\chi_0^{-1}$, Fig.~\ref{fig:2b}(c), leading to a step in $N_-$, Fig.~\ref{fig:2b}(d), but otherwise has little effect.

Once $N_-$ has been calculated, Eq.~\ref{eq:5} can be
evaluated numerically.  The singular part of the integral is treated analytically, and the remainder numerically, with  
$u$ and $U$ approximated as constants, $u=0.8eV^{-1}$, while $U$ = 2~eV 
unless otherwise noted.
%at half filling (Fig.~3), 0.83~eV for the hole doped states of Fig.~4.  This doping dependence of 
%$U$ is consistent with earlier results\cite{kusko,tanmoyop}.  
It is convenient to fix the SDOS at some temperature $T'$, and then solve Eq.~\ref{eq:5} for $T(\xi_{th} ,T')$, with self-consistency requiring $T(\xi_{th} ,T')=T'$.  %Figure~\ref{fig:3g}(b) shows the calculated correlation length for this LSCO model (red solid line with filled dots), compared to a reference model displaying conventional HM behavior (blue dashed line with triangles).  The reference data are generated by taking the $T=4000$K SDOS at all $T$, but stretching and shifting it to lie in the same $X_-$-range as the lower $T$ SDOS.  The resulting $\xi_{th}$ grows smoothly, following Eq.~\ref{eq:4}.  
 
%It is not a specifically 2D effect, but should also arise in 3D materials.

When the OZ form of $\chi$ is assumed, $N_-$ becomes a constant, which we denote $N_a$, and Eq.~\ref{eq:5} leads to long range order at $T=0$ only, with correlation length $\xi_{th}/a=1/\sqrt{4\pi N_a\delta}$ given by
\begin{eqnarray}
\xi_{th} q_c=e^{T_2/T},
\label{eq:4}
\end{eqnarray}
where $\delta = U_c-U_{sp}$, $U_c= 1/\chi_0({\bf Q_0},0)$, 
$T_2=\pi A_2\lambda /6\Gamma ua^2$, $A_i$ is the coefficient of $q^i$, and 
$q_c$ is a wave number cutoff.  

The origin of the anomalous region II for undoped LSCO, Fig.~\ref{fig:3g}(b), can be readily understood from Fig.~\ref{fig:3g}(a), where for $T<1500K$ there is excess SDOS weight near $U_c$, leading to a strong peak (inset) as $T\rightarrow 0$.  This feature represents the development of the $(\pi,\pi)$-plateau.   For the small-$t'$ materials, the susceptibility on the plateau remains parabolic, but with a small curvature and a sharp cutoff.  This leads to an additional contribution to the SDOS of the form $N_-=N_p$ if $X_-\le  X_p$.
%To approximately account for this, we can add a term $c(T)\delta(X_--U_c)$ to the SDOS, where $c(T)$ is the excess height of $N$ at threshold, proportional to the area of the plateau at $T=0$.  
Then the integral of Eq.~\ref{eq:5} becomes
\begin{eqnarray}
I=(N_a-N_b\delta)ln(\frac{X_c}{\delta}+1)+N_bX_c+N_p ln(\frac{X_p}{\delta}+1). %\\
%\nonumber
%\rightarrow aln(\frac{X_c}{\delta})+bX_c+\frac{c}{\delta},
\label{eq:4b}
\end{eqnarray}
where $X_c$ is the maximum of $X_-$.  The last term, which we denote $I_p$, has two distinct limits.  If $X_p<<\delta$, then $I_p\rightarrow N_pX_p/\delta$.  This is equivalent to assuming that the plateau has a flat top, in which case $N_-$ can be represented as a $\delta$-function, $N_-= N_pX_p\delta(X_--U_c)$, with $N_pX_p$ the excess height of $N$ at threshold, proportional to the area of the plateau at $T=0$.  As $T$ decreases, $\delta$ also decreases, and the opposite limit for $I_p$ becomes appropriate, $\delta<<X_p$, in which case $I_p=N_pln(X_p/\delta )$.  From Eq.~\ref{eq:5}, $I_p\sim 1/\delta$ translates into $\xi_{th}\sim 1/T^{1/2}$, dotted line in Fig.~\ref{fig:3g}(b).  The actual anomaly roughly follows this line, but the correlation length growth is actually reversed.
%power-law scaling regimes can be derived analytically.  For LSCO at $x=0$, in , the this effect can be understood 

The present $G_Z-W_Z-\Gamma_Z$ model is the simplest which captures the essential physics of the pseudogap.  For a more quantitative comparison with experiment, two additional problems must be solved.  First, an extension to fully self-consistent $G-W-\Gamma$ may be needed to capture the splitting of the saddle VHS peak into the VHS exciton and the residual continuum part.  Secondly, the term in Eq.~\ref{eq:3} proportional to $\omega^z$ must be included to describe quantum fluctuations.  Since we are here primarily interested in the opening of the pseudogap at higher temperatures, we ignored it in our analysis.  Finally, it will be interesting to explore if excitonic superfluidity exists, and how it is related to high-$T_c$ superconductivity.
%An additional shortcoming of the OZ analysis is a lack of self-consistency.  The OZ parameters are calculated from the susceptibility of the paramagnetic phase, whereas a susceptibility of the dressed bands is more appropriate deep in the pseudogap state.  Again, since we are mainly interested in higher-$T$ effects, we will adopt this simplifying assumption, but will point out where it may limit our analysis.

\section{References}
\begin{enumerate}

\item Kivelson, S.A. Bindloss, I.P. Fradkin, E. Oganesyan, V. Tranquada, J.M. Kapitulnik, A. \& Howald, C. 
How to detect fluctuating order in the high-temperature superconductors.
{\it Rev. Mod. Phys.} {\bf 75,} 1201-1241 (2003).
\item Vojta, M. 
Lattice symmetry breaking in cuprate superconductors: stripes, nematics, and superconductivity.
{\it Adv. Phys.} {\bf 58,} 699-820 (2009).

\item Wu, T. et al.
%H. Mayaffre, S. Kr\"amer, M. Horvati\'c, C. Berthier, W.N. Hardy, Ruixing Liang, D.A. Bonn, and M.-H. Julien, 
Magnetic-field-induced charge-stripe order in the high-temperature superconductor YBa$_2$Cu$_3$O$_y$.
{\it Nature} {\bf 477,} 191-194 (2011).
\item Ghiringhelli, G. et al.
%M. Le Tacon, M. Minola, S. Blanco-Canosa, C. Mazzoli, N. B. Brookes, G.M. De Luca, A. Frano, D.G. Hawthorn, F. He, T. Loew, M. Moretti Sala, D.C. Peets, M. Salluzzo, E. Schierle, R. Sutarto, G.A. Sawatzky, E. Weschke, B. Keimer, and L. Braicovich, 
Long-range incommensurate charge fluctuations in (Y,Nd)Ba$_2$Cu$_3$O$_{6+x}$.
{\it Science} {\bf 337,} 821-825 (2012). 
\item Achkar, A.J. et al.
%R. Sutarto, X. Mao, F. He, A. Frano, S. Blanco-Canosa, M. Le Tacon, G. Ghiringhelli, L. Braicovich, M. Minola, M. Moretti Sala, C. Mazzoli, Ruixing Liang, D.A. Bonn, W.N. Hardy, B. Keimer, G.A. Sawatzky, and D.G. Hawthorn, 
Distinct charge orders in the planes and chains of ortho-III-ordered YBa$_2$Cu$_3$O$_{6+\delta}$ superconductors identified by resonant elastic x-ray scattering.
{\it Phys. Rev. Lett.} {\bf 109,} 167001 (2012).
\item Chang, J. et al.
%E. Blackburn, A.T. Holmes, N.B. Christensen, J. Larsen, J. Mesot, Ruixing Liang, D.A. Bonn, W.N. Hardy, A. Watenphul, M. v. Zimmermann, E.M. Forgan, and S.M. Hayden, 
Direct observation of competition between superconductivity and charge density wave order in YBa$_2$Cu$_3$O$_{6.67}$.
{\it Nature Phys.} {\bf 8,} 871-876 (2012). %arXiv:1206.4333.
\item  LeBoeuf, D. et al.
%S. Kr\"amer, W.N. Hardy, Ruixing Liang, D.A. Bonn, and C. Proust, 
Thermodynamic phase diagram of static charge order in underdoped YBa$_2$Cu$_3$O$_y$.
{\it Nature Phys.} {\bf 9,} 79-83 (2013). %arXiv:1211.2724.
\item  Blackburn, E. et al. 
%J. Chang, M. Hucker, A.T. Holmes, N.B. Christensen, Ruixing Liang, D.A. Bonn, W N. Hardy, M. v. Zimmermann, E.M. Forgan, and S.M. Hayden, 
X-say diffraction observations of a charge-density-wave order in superconducting ortho-II YBa$_2$Cu$_3$O$_{6.54}$ single crystals in zero magnetic field.
{\it Phys. Rev. Lett.} {\bf 110,} 137004 (2013). 
%\item{footx}A second possible route to nematic coupling is discussed in the Supplementary Material.
\item Doiron-Leyraud, N. et al.
%S. Lepault, O. Cyr-Choiniere, B. Vignolle, G. Grissonnanche, F. Laliberte, J. Chang, N. Barisic, M. K. Chan, L. Ji, X. Zhao, Y. Li, M. Greven, C. Proust, L. Taillefer, 
Hall, Seebeck, and Nernst coefficients of underdoped HgBa$_2$CuO$_{4+\delta}$: Fermi-surface reconstruction in an archetypal cuprate superconductor.
{\it Phys. Rev. X}{\bf 3,} 021019 (2013). 
\item Comin, R. et al.
%A. Frano, M.M. Yee, Y. Yoshida, H. Eisaki, E. Schierle, E. Weschke, R. Sutarto, F. He, A. Soumyanarayanan, Yang He, M. Le Tacon, I.S. Elfimov, Jennifer E. Hoffman, G.A. Sawatzky, B. Keimer, and A. Damascelli,
Charge order driven by Fermi-arc instability in Bi$_2$Sr$_{2−x}$La$_x$CuO$_{6+\delta}$.
{\it Science} {\bf 343,} 390-392 (2014).
\item da Silva Neto, E.H. 
%Pegor Aynajian, Alex Frano, Riccardo Comin, Enrico Schierle, Eugen Weschke, Andr\'as Gyenis, Jinsheng Wen, John Schneeloch, Zhijun Xu, Shimpei Ono, Genda Gu, Mathieu Le Tacon, and Ali Yazdani, 
Ubiquitous interplay between charge ordering and high-temperature superconductivity in cuprates.
{\it Science} {\bf 343,} 393-396 (2014). 
\item Fujita, K. et al.
%C.K. Kim, I. Lee, J. Lee, M.H. Hamidian, I.A. Firmo, S. Mukhopadhyay, H. Eisaki, S. Uchida, M.J. Lawler, E.-A. Kim, and J C. Davis,  
Simultaneous transitions in cuprate momentum-space topology and electronic symmetry breaking.
{\it Science} {\bf 344,} 612-616 (2014).  %arXiv:1403.7788.
\item Metlitski, M.A. \& Sachdev, S. 
Quantum phase transitions of metals in two spatial dimensions. II. Spin density wave order.
{\it Phys. Rev. B}{\bf 82,} 075128 (2010).

\item Wang, Y.  \& Chubukov, A.V. 
Charge-density-wave order with momentum (2Q,0) and (0,2Q) within the spin-fermion model: continuous and discrete symmetry breaking, preemptive composite order, and relation to pseudogap in hole-doped cuprates.
{\it Phys. Rev. B}{\bf 90,} 035149 (2014).
\item Efetov, K.B. Meier, H. \& P\'epin, C. 
Pseudogap state near a quantum critical point.
{\it Nature Phys.} {\bf 9,} 442-446 (2013).
\item Meier, H. P\'epin, C. Einenkel, M. \& Efetov, K.B. 
Cascade of phase transitions in the vicinity of a quantum critical point.
{\it Phys. Rev. B}{\bf 89,} 195115 (2014). %arxiv:1312.2010.
\item La Placa, R. \& Sachdev, S. 
Bond order in two-dimensional metals with antiferromagnetic exchange interactions.
{\it Phys. Rev. Lett.} {\bf 111,} 027202 (2013).
\item Hayward, L.E. Hawthorn, D.G. Melko, R.G. \& Sachdev, S. 
Angular fluctuations of a multicomponent order describe the pseudogap of YBa$_2$Cu$_3$O$_{6+x}$.
{\it Science} {\bf 343,} 1336-1339 (2014). %arxiv:1309.6639.
\item Bulut, S. Atkinson, W.A. \& Kampf, A.P. 
Spatially modulated electronic nematicity in the three-band model of cuprate superconductors.
{\it Phys. Rev. B}{\bf 88,} 155132 (2013).
\item Allais, A. Bauer, J. \& Sachdev, S. 
Density wave instabilities in a correlated two-dimensional metal.
{\it Phys. Rev. B}{\bf 90,} 155114 (2014). % arXiv:1402.4807,6311.
\item Fujita, K. et al. 
%M.H. Hamidian, S.D. Edkins, C.K. Kim, Y. Kohsaka, M. Azuma, M. Takano, H. Takagi, H. Eisaki, S. Uchida, A. Allais, M.J. Lawler, E.-A. Kim, S. Sachdev, and J.C. S\'eamus Davis, 
Direct phase-sensitive identification of a d-form factor density wave in underdoped cuprates.
{\it Proc. Nat. Acad. Sci. of the USA} {\bf111,} E3026-E3032 (2014).
\item Markiewicz, R.S.  Lorenzana, J. Seibold, G. \& Bansil, A. 
Gutzwiller charge phase diagram of cuprates, including electron-phonon coupling effects.
{\it New Journal of Physics} {\bf 17,} 023074 (2015).
%Short range smectic and long range nematic order in the pseudogap phase of cuprates.
%{\it arXiv}:1207.5715.

\item Hertz, J.A. 
Quantum critical phenomena.
{\it Phys. Rev. B}{\bf 14,} 1165-1184 (1976).
\item Millis, A.J. 
Effect of a nonzero temperature on quantum critical points in itinerant fermion systems.
{\it Phys. Rev. B}{\bf 48,} 7183-7196 (1993).

\item Ornstein, L. S. and Zernike, F. 
Accidental deviations of density and opalescence at the critical point of a single substance. 
{\it Proc. Acad. Sci. Amsterdam} {\bf  17,} 793-806 (1914).
\item Overhauser, A.W. 
Exchange and correlation instabilities of simple metals.
{\it Phys. Rev.} {\bf 167,} 691-698 (1968).
%, Phys. Rev. B{\bf 2}, 874 (1970).

\item McMillan, W.L. 
Microscopic model of charge-density waves in 2H−TaSe$_2$.
{\it Phys. Rev. B}{\bf 16,} 643-650 (1977).
\item Motizuki, K. \& Suzuki, N. 
{\it Structural Phase Transitions in Layered Transition-Metal Compounds} (Reidel, Dordrecht, 1986).
\item Yoshiyama, H. Takaoka, Y. Suzuki, N. \& Motizuki, K. 
Effects on lattice fluctuations on the charge-density-wave transition in transition-metal dichalcogenides.
{\it J. Phys. C} {\bf 19,} 5591-5606 (1986).
\item Halperin, B.I. \& Rice, T.M.
The excitonic state at the semiconductor-semimetal transition.
in {\it Solid State Physics}, Vol.~21, ed. Seitz, F. Turnbull, D. \& Ehrenreich H.
(New York, Academic) pp. 115-192.
\item Bronold, F.X., \& Fehske, H.
Possibility of an excitonic insulator at the semiconductor-semimetal transition. 
{\it Phys. Rev. B}{\bf 74,} 165107 (2006).
\item C\^ot\'e, R., \& Griffin, A.
Excitonic modes in a Bose-condensed electron-hole gas in the pairing approximation.
{\it Phys. Rev. B}{\bf 37,} 4539-4551 (1988).
\item Markiewicz, R.S. 
Mode-coupling model of Mott gap collapse in the cuprates: Natural phase boundary for quantum critical points
{\it Phys. Rev. B}{\bf 70,} 174518 (2004). 
\item Markiewicz, R.S. Lorenzana, J. Seibold, G. \& Bansil,  A.
Gutzwiller magnetic phase diagram of the cuprates.
{\it Phys. Rev. B}{\bf 81,} 014509 (2010).
\item Phillips, J.C.
Ultraviolet absrption of insulators III: fcc alkali halides
{\it Phys. Rev.} {\bf 136,} A1705 (1964).
%\item Markiewicz, R.S. Sahrakorpi, S. \& Bansil, A.
%Paramagnon-induced dispersion anomalies in the cuprates.
%{\it Phys. Rev. B}{\bf 76,} 174514 (2007).

\item Andergassen, S. Caprara, S. Di Castro, C. \& Grilli, M. 
Anomalous isotopic effect near the charge-ordering quantum criticality.
{\it Phys. Rev. Lett.} {\bf 87,} 056401 (2001).
\item  Nagaosa, N.
{\it Quantum Field Theory in Condensed Matter Physics}, (Berlin, Springer, 1999).
\item Yamada, K. 
{\it Electron Correlation in Metals}, (Cambridge, University Press, 2004).
\item Mermin N.D. \& Wagner, H. 
Absence of ferromagnetism or antiferromagnetism in one- or two-dimensional isotropic Heisenberg models.
{\it Phys. Rev. Lett.} {\bf 17,} 1133-1136 (1966).

\item Birgeneau, R.J. et al.
%A. Aharony, N.R. Belk, F.C. Chou, Y. Endoh, M. Greven, S. Hosoya, M.A. Kastner, C.H. Lee, Y.S. Lee, G. Shirane, S. Wakimoto, B.O. Wells, and K. Yamada, 
Magnetism and magnetic fluctuations in La$_{2-x}$Sr$_x$CuO$_4$ for x=0 (2D antiferromagnet), 0.04 (3D spin glass) and x=0.15 (superconductor).
{\it J. Phys. Chem. Solids} {\bf 56,} 1913-1920 (1995).
%\item }L.F. Tocchio, F. Becca, A. Parola, and S. Sorella, Phys. Rev. B 78 (2008), p. 041101(R).
\item Kusko, C. Markiewicz, R.S. Lindroos, M. \& Bansil, A. 
Fermi surface evolution and collapse of the Mott pseudogap in Nd$_{2−x}$Ce$_x$CuO$_{4±\delta}$.
{\it Phys. Rev. B}{\bf 66,} 140513(R) (2002).
\item Das, T. Markiewicz, R.S. \&  Bansil, A.
Optical model-solution to the competition between a pseudogap phase and a charge-transfer-gap phase in high-temperature cuprate superconductors.
{\it Phys. Rev. B}{\bf 81,} 174504 (2010).

%\item Markiewicz, R.S. 
%Van Hove excitons and High-T$_c$ superconductivity
%(IV) Properties of the excitons.
%{\it Physica C} {\bf 168,} 195-204 (1990).
%%Excitons at a Van Hove singularity
%%{\it J. Phys.: Cond. Matt.} {\bf 3,} 3859-3863 (1991).
%\item{ExIns2}Onufrieva, F. \& Pfeuty, P.
%Quantum critical point associated with the electronic topological transition in a two-dimensional electron system as a driving force for anomalies in underdoped high-Tc cuprates.
%{\it Phys. Rev.}{\bf B 61,} 799-820 (2000). 

\item Piriou, A. Jenkins, N. Berthod, C. Maggio-Aprile, I. \&  Fischer, \O. 
First direct observation of the Van Hove singularity in the tunnelling spectra of cuprates.
{\it Nature Communications} {\bf 2,} 221 (2011). %DOI:10.1038/ncomms1229.
\item Nieminen, J. Suominen,  I. Das, T.  Markiewicz, R.S.\& Bansil, A. 
Evidence of strong correlations at the van Hove singularity in the scanning tunneling spectra of superconducting Bi$_2$Sr$_2$CaCu$_2$O$_{8+\delta}$ single crystals.
{\it Phys. Rev. B}{\bf 85,} 214504 (2012).
\item Benhabib, S. et al.
%A. Sacuto, M. Civelli, I. Paul, M. Cazayous, Y. Gallais, M.-A. Measson, R. D. Zhong, J. Schneeloch, G. D. Gu, D. Colson, and A.Forget, 
Collapse of the normal state pseudogap at a Lifshitz transition in Bi$_2$Sr$_2$CaCu$_2$O$_{8+\delta}$ cuprate superconductor.
{\it Phys. Rev. Lett.} {\bf 114,}147001 (2015).%arXiv:1403.7620}.
\item Buhmann, J.M. Ossadnik, M. Rice, T.M. \& Sigrist, M. 
Numerical study of charge transport of overdoped La$_{2−x}$Sr$_x$CuO$_4$ within semiclassical Boltzmann transport theory.
{\it Phys. Rev. B}{\bf 87,} 035129 (2013).
\item Barzykin, V. \& Pines, D. 
Universal behavior and a two-fluid description of the cuprate superconductors.
{\it Adv. Phys.} {\bf 58,} 1-65 (2009).
\item Curro, N. Fisk, Z. \& Pines, D.
Scaling and the magnetic origin of emergent behavior in correlated electron superconductors.
{\it MRS Bulletin} {\bf 30,} 442-446 (2005).
\item Aeppli, G. Mason, T.E. Hayden, S.M. Mook,  H.A. \&  Kulda, J.
Nearly singular magnetic fluctuations in the normal state of a high-T$_c$ cuprate superconductor
{\it Science} {\bf 278,}1432-1435 (1997). 
\item Das, T. Markiewicz, R.S. \& Bansil, A. 
Intermediate coupling model of the cuprates.
{\it Advances in Physics} {\bf 63,} 151-266 (2014).
\item Vilk, Y.M. \& Tremblay, A.-M.S. 
Non-perturbative many-body approach to the Hubbard model and single-particle pseudogap.
{\it J. Phys. I France} {\bf 7,} 1309-1368 (1997).
\item Senthil, T. Vishwanath, A. Balents, L. Sachdev, S. \&  Fisher, M.P.A.
Deconfined quantum critical points
{\it Science }{\bf 303,} 1490-1494 (2004).
\item  Chen,  K.-S. Meng, Z.Y. Pruschke, T. Moreno, J. \& Jarrell, M.
Lifshitz transition in the two-dimensional Hubbard model
{\it Phys. Rev. B}{\bf 86,} 165136 (2012).
\item  Rice, T.M. Yang, K.-Y. \& Zhang, F.C. 
A phenomenological theory of the anomalous pseudogap phase in underdoped cuprates.
{\it Rep. Prog. Phys.} {\bf 75,} 016502 (2012).
\item Rice T.M. \& Scott, G.K. 
New mechanism for a charge-density-wave instability
{\it Phys. Rev. Lett.} {\bf 35,} 120-123 (1975).
\item Budai, J.D. et al.
%Jiawang Hong, Michael E. Manley, Eliot D. Specht, Chen W. Li, Jonathan Z. Tischler, Douglas L. Abernathy, Ayman H. Said, Bogdan M. Leu, Lynn A. Boatner, Robert J. McQueeney & Olivier Delaire
Metallization of vanadium dioxide driven by large phonon entropy
{\it Nature} {\bf 515,} 535–539 (2014).
\item Bechstedt, F.,
{\it Many-Body Approach to Electronic Excitations}
(Springer, Berlin, 2015).
\end{enumerate}
\section{Acknowledgements}     
%{\bf Acknowledgments} This work is supported by the US Department of Energy, Office of Science, Basic Energy Sciences contract DE-FG02-07ER46352, and benefited from the allocation of supercomputer time at NERSC and Northeastern University's Advanced Scientific Computation Center (ASCC).  
This work is supported by the US Department of Energy, Office of Science, Basic Energy Sciences grant number DE-FG02-07ER46352, and benefited from Northeastern University's Advanced Scientific Computation Center (ASCC) 
%theory support at the Advanced Light Source, Berkeley 
and the allocation of supercomputer time at NERSC through grant number DE-AC02-05CH11231.
\section{Author contributions}
R.S.M., I.G.B., P.M., and A.B. contributed to the research reported in this study and the writing of the manuscript.
\section{Additional information}
The authors declare no competing financial interests. 
\end{document}

%****
\title{{\color{red}SUPPLEMENTARY MATERIAL for}\\
Condensation Bottleneck Driven by a Hidden Van Hove Singularity as the  Origin of Pseudogap Physics in Cuprate High-$T_c$ Superconductors}
\author{R.S. Markiewicz}
%\affiliation{ Physics Department, Northeastern University, Boston MA 02115, USA}
\author{I.G. Buda}
%\affiliation{ Physics Department, Northeastern University, Boston MA 02115, USA}
\author{P. Mistark}
%\affiliation{ Physics Department, Northeastern University, Boston MA 02115, USA}
%\author{J. Lorenzana}
%\affiliation{ISC-CNR and Dipartimento di Fisica, Universit\`a di Roma ``La Sapienza'', P. Aldo Moro 2, 00185 Roma, Italy}
%\affiliation{ISC-CNR, Via dei Taurini 19, I-00185 Roma, Italy}
%\author{G. Seibold}
%\affiliation{Institut F\"ur Physik, BTU Cottbus-Senftenberg, PBox 101344, 03013 Cottbus,
%Germany} 
\author{A. Bansil}
%\affiliation{ Physics Department, Northeastern University, Boston MA 02115, USA}

\maketitle

%\appendix

\section{Genetic algorithm for Cuprates}
\subsection{A note on normalization conventions}

In this paper susceptibilities are calculated at intermediate coupling level, using $Z$-corrected first-principles dispersions to describe the coherent part of the quasiparticle-GW dressed carriers.\cite{AIP}  The results can be expressed in two different ways.  If the susceptibility $\chi_0$ is calculated based on the density functional theory (DFT) 
%local-density approximation (LDA) 
dispersion $\epsilon_0(k)$, then the phase diagrams are expressed in terms of renormalized temperatures $T_Z=T/Z$ and Hubbard $U$'s, $U_Z=ZU$, where $Z$ is the QPGW renormalization factor.  This is used in the Supplementary Material.  For convenience in comparing with experiment, in the main text we calculated $\chi$ based on the dressed dispersion $\epsilon_Z=Z\epsilon_0$, assuming a doping-independent $Z=0.5$.  

Also, when referring to Figures or Equations in the main text, we will use a leading `M', as in Fig. M1.

\subsection{Cuprate algorithm}

\begin{figure}
\leavevmode  
\rotatebox{0}{\scalebox{0.4}{\includegraphics{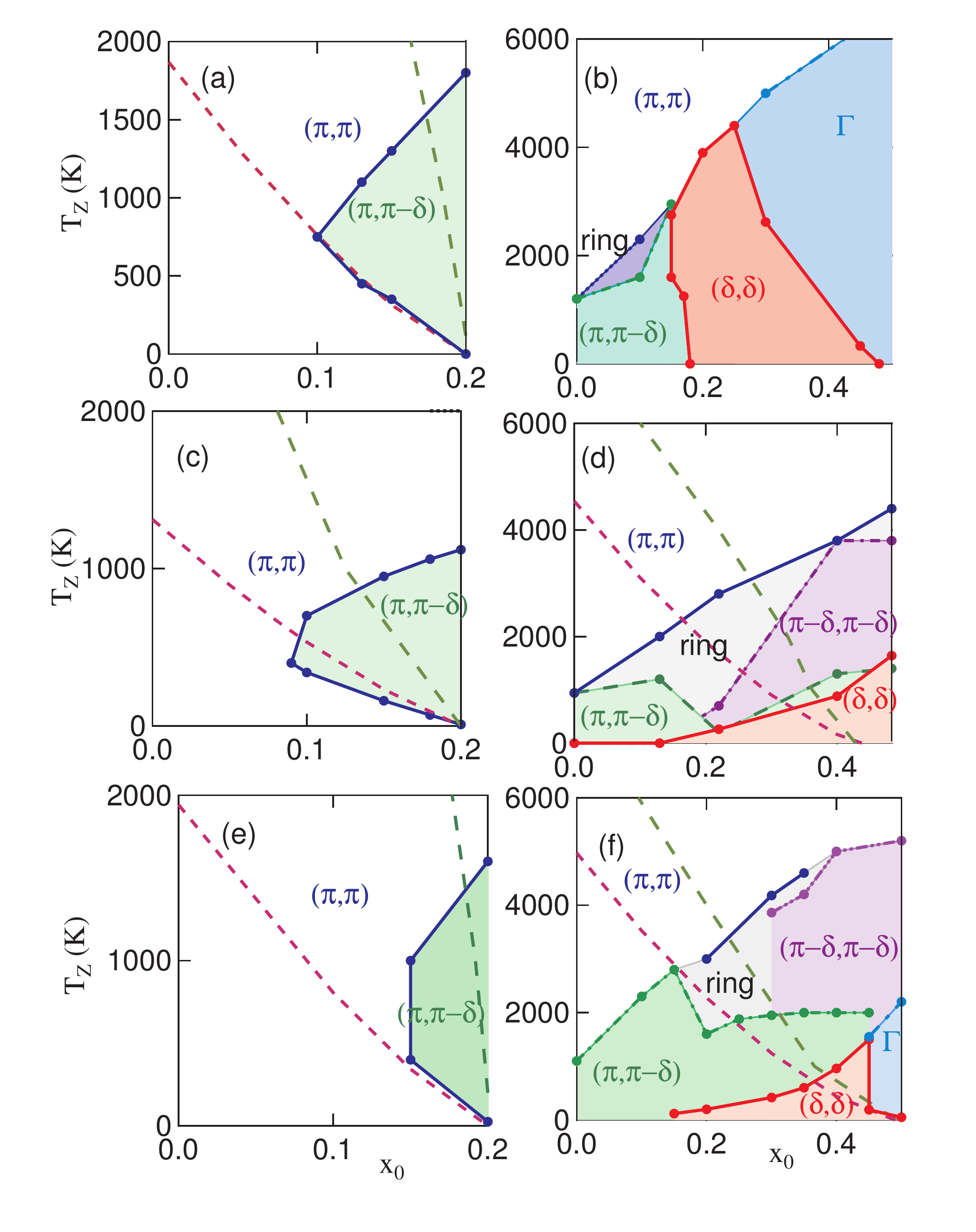}}}
%\vskip0.5cm
\caption{%(Color online.)  
{\bf Fluctuation fingerprints of cuprates.}
Maps of $q$-vectors of largest susceptibility for DFT-based model of LSCO (c) and Bi2201 (d) compared to $t-t'-t''$ reference phase a [$t''$ = 0] with $t'$ = -0.23$t$ (a) or -0.43$t$ (b), and reference phase b [$t''=-0.5t'$] with $t'$ = -0.12$t$ (e) or -0.258$t$ (f). Recall, from Fig.~M1(c), that $x_0$ is the doping at $T=0$, but we assume $x\simeq x_0$. }
\label{fig:009} 
\end{figure}

To understand differences between cuprates, and to gauge their proximity to the pure Hubbard model, it is necessary to add a third axis to the usual $T$- and doping-phase diagrams.  Here we introduce a technique to reduce differences in band structure to the variation of a single hopping parameter.  

While DFT has been liberating in that it provides a parameter-free starting point for studies of materials, it can also be constraining, in that each material is defined as a single point in material parameter space, and it is hard to define connections between the members of a family of materials, such as cuprates or pnictides.  What is needed is a form of materials genome, with an algorithm defining the relatedness of materials, the evolution from one material to another, and the probable route for extending the family to new materials.  Here we propose such an algorithm for the cuprates -- a `cuprate genome'.  The cuprates are particularly simple, in that their properties are dominated by a single CuO$_2$ band crossing the Fermi level, so their parameter space consists of hopping parameters, traditionally $t$, $t'$, $t''$, ..., which can be determined, e.g., by a Wannier-type downfolding of the DFT bands.  In fact, since $t$ sets an energy scale, the band dispersion is determined by the ratios $t'/t$, $t''/t$, ... In genetics, a family would be defined as a group lying on a low-dimensional surface in parameter space.  The simplest genetics is associated with changes of a single gene, called Mendelian.  Similarly, the simplest material family would lie along a single line in material space.  We show that the cuprates correspond to such a Mendelian family, by introducing {\it reference families} depending only on $t'/t$, $t''/t$, and show that the cuprates lie along a particular cut in this 2D-space.

To do this, we introduce a definition of equivalence: ideally, two materials are equivalent if they have the same phase diagram.  However, the real phase diagrams are defined by interactions beyond the current scope of DFT theories.  The best we can do is to say that they would have the same phase diagrams in the context of DFT.  For weak and intermediately correlated materials, the ordered phases are generally determined from the Stoner criterion, and hence correlated with peaks in the Lindhard susceptibility at $\omega =0$, which can readily be calculated.

Figure~\ref{fig:009} is a map of the $q$-vectors associated with the largest susceptibilities, calculated for a variety of dispersions.  The left-hand frames (a), (c), (e), refer to LSCO, the right-hand frames (b), (d), (f), to Bi2201.  The middle row (c), (d), uses the full DFT dispersion, compared to two reference families, with $(a)$ cuts defined by $t''=0$ (top row, (a), (b)) or $(b)$
 the Pavarini-Andersen cut\cite{PavOK}, $t''=-t'/2$ (bottom row, (e), (f)).  We find that reference family $b$ semi-quantitatively reproduces the phase diagram for both cuprates.  On the other hand, setting $t''=0$ works well for LSCO, Fig.~\ref{fig:009}(a), but fails for Bi2201  Fig.~\ref{fig:009}(b).   We further find that a reasonable choice of $t'/t$ is the one which matches the Van Hove singularity (VHS) doping of the full DFT dispersion.  In these figures we fix an energy scale by taking $t$ = 0.4195~eV, appropriate for LSCO\cite{RSM_tb}.   We note that both reference families have the additional desirable feature that they evolve from the state with $t'=0$ -- i.e., the original Hubbard model.

This confirms our earlier finding that LSCO, Fig.~\ref{fig:009}(c), has a very different phase diagram from most other cuprates, here represented by Bi2201, Fig.~\ref{fig:009}(d).\cite{Gzm1}  In the phase labeled `ring', the susceptibility is approximately constant along a ring of $q$-vectors surrounding $(\pi,\pi)$.  We note in particular that antinodal nesting (red-shaded region) emerges naturally in most cuprates at higher doping, but is absent in LSCO.  Shown also in Fig.~\ref{fig:009} are $T_{VHS}$ (dark red dashed line) and the shift of the DOS-peak (thin green dashed line), discussed in the main text for LSCO.

\begin{figure}
\leavevmode  
\rotatebox{0}{\scalebox{0.35}{\includegraphics{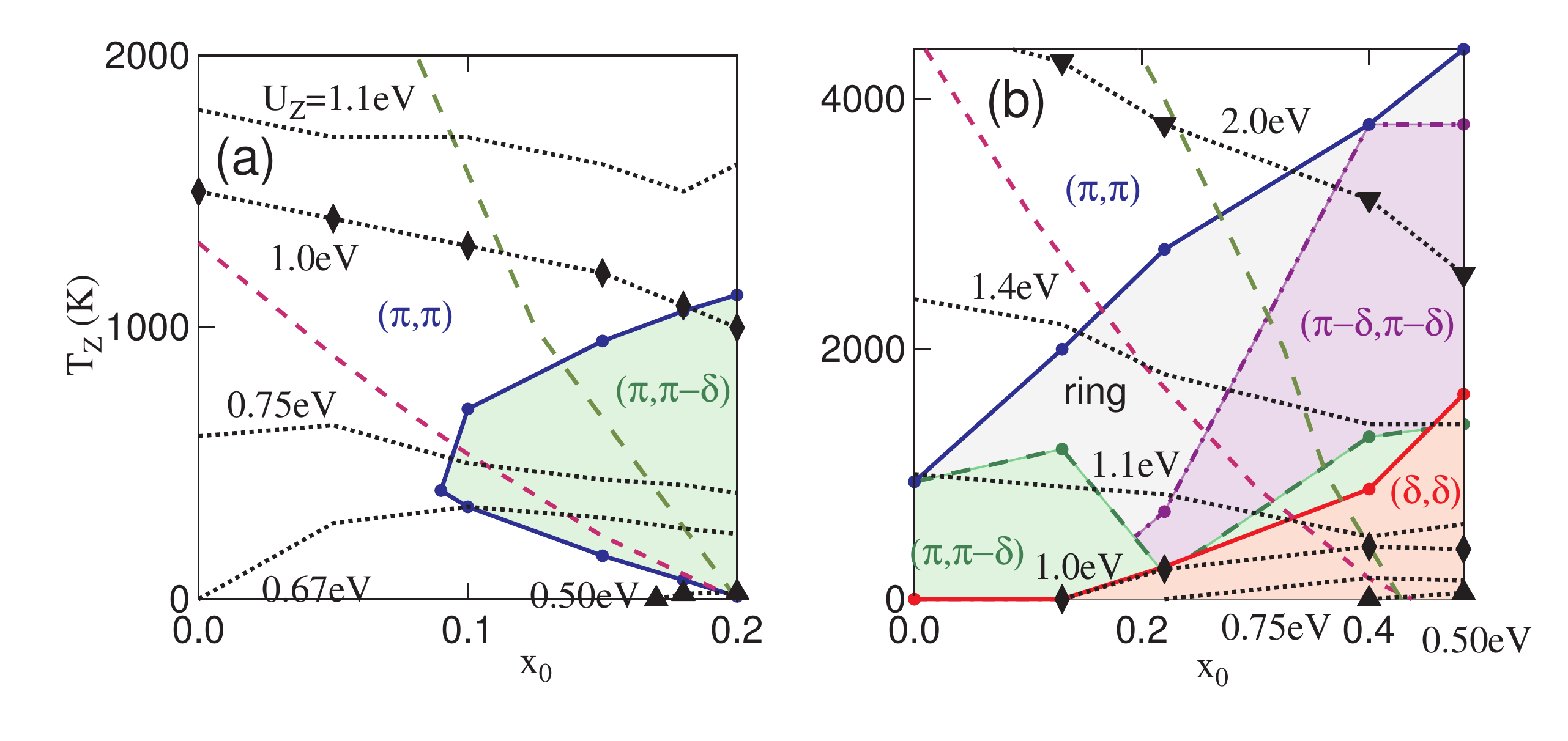}}}
%\vskip0.5cm
\caption{%(Color online.)  
{\bf Mean-field phase diagrams of cuprates}, for DFT-based models of LSCO (a) and Bi2201 (b).  Black dots indicate mean field phase boundaries for different $U_Z$ values, including $U_Z$ = 2 (down triangles), 1 (diamonds), or 0.5~eV (up triangles).
} 
\label{fig:009c} 
\end{figure}
In Fig.~\ref{fig:009c},  we replot the DFT fluctuation maps, adding the mean-field phase boundaries, based on a Stoner criterion, 
\begin{equation}
U_Z\chi_0(q,\omega=0)=1.
\label{eq:0aa}
\end{equation}
%where the  bare Lindhard susceptibility $\chi_0(q,\omega)$ is based on the DFT bands, $U_Z=ZU$ is a renormalized Hubbard interaction\cite{Gzm1,AIP}, $Z$ is the self-consistent quasiparticle-GW (QP-GW) renormalization factor, and $U$ is the bare Hubbard $U$.   
In all cases, there is a critical value of $U=U_0$ below which there is no phase transition at $x=0$.  For $U<U_0$, the phase boundary is dome-shaped, but with a peak shifted away from the $T=0$ value of $x_{VHS}$ to a value close to the line $x_{VHS}(T)$, red-dashed line.  For $U>U_0$ the phase boundary rapidly evolves to a strong coupling form where the highest $T$ transition occurs near half filling.  This is consistent with our finding in the main text that there is an apparent shift of the VHS toward $x=0$ as $T$ increases. In all cases, at sufficiently high $T$ FS nesting is washed out and the dominant instability is at $q=(\pi,\pi)$.  

\begin{figure}
\leavevmode  
\rotatebox{0}{\scalebox{0.45}{\includegraphics{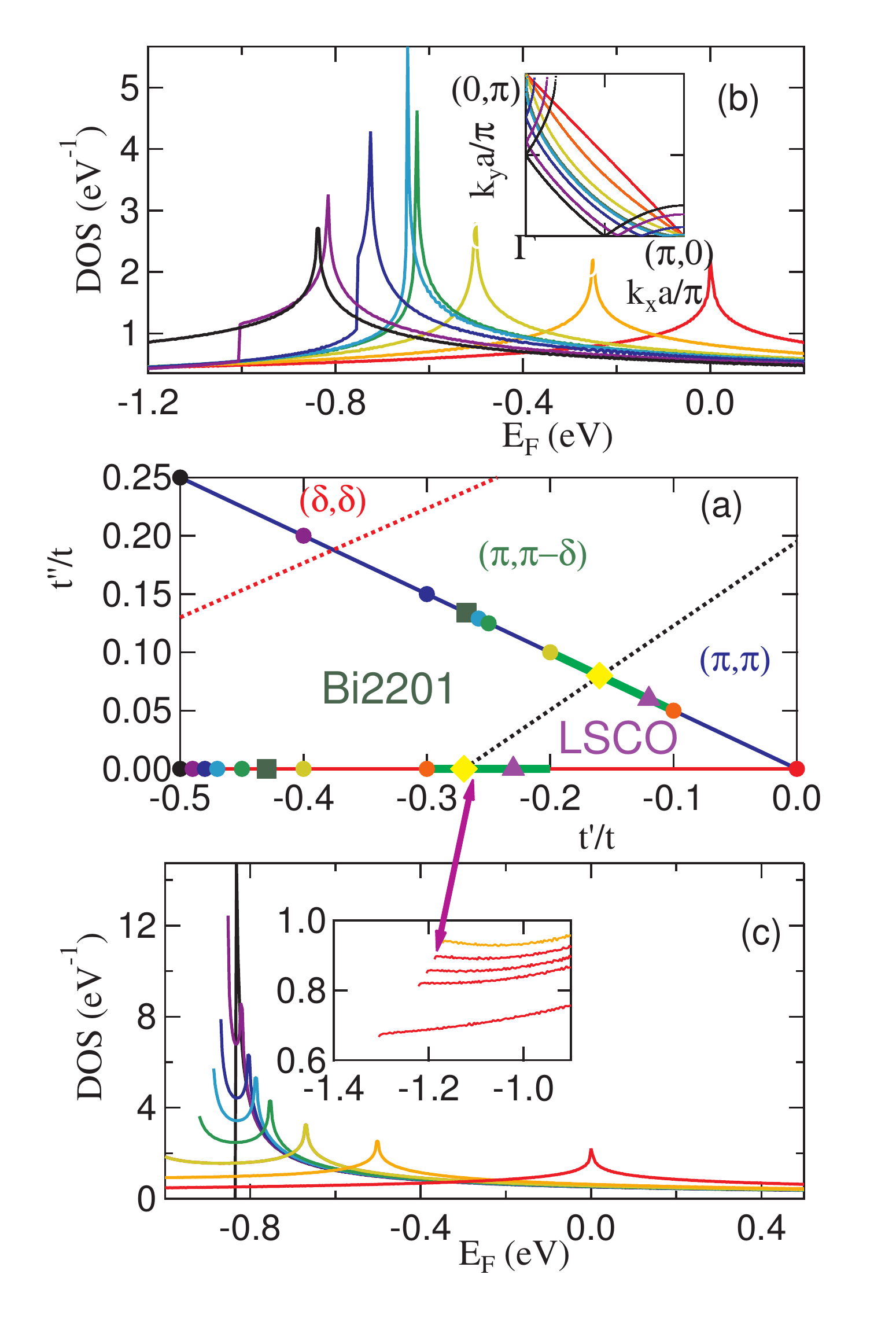}}}
%\vskip0.5cm
\caption{%(Color online.)  
{\bf Reference families of the cuprates.} (a) Reference cuts in $t'-t''$-space.  Dotted lines indicate ground state transitions at $x=0$.  Light green line segments indicate the ranges over which $\xi$ is significantly depressed, Fig.~M2, with maximal depression indicated by yellow diamonds.  Violet triangles [green squares] = reference states for LSCO [Bi2201]; colored circles indicate $t'$-values for the corresponding DOSs in frame (b) [upper line] or (c) [lower line].  (b)  DOS for several values of $t'$ for reference family $b$ ($t''=-t'/2$).   As VHS moves from right to left, $t'/t$ = 0 (red curve), -0.1 (orange), -0.2 (yellow-green), -0.25 (green), -0.258 (light blue), -0.3 (blue), -0.4 (violet),  and -0.5 (black).   Inset: corresponding Fermi surfaces at VHS.  (c)  DOS for several values of $t'$ for reference family $a$ ($t''=0$).  As VHS moves from right to left, $t'/t$ = 0 (red curve), -0.3 (orange), -0.4 (yellow-green), -0.45 (green), -0.47 (light blue), -0.48 (blue), -0.49 (violet), and -0.5 (black). Inset: leading edge of DOS, near band bottom for (from bottom to top) $t'/t$ = -0.22, -0.27, -0.28, -0.29, -0.30, in regime where it crosses over from a local minimum to a local maximum.
} 
\label{fig:009d} 
\end{figure}
An advantage of using the genetic algorithm is that it lets us systematically vary parameter space, to study the evolution from one cuprate to another, as in Fig.~M2, or to extrapolate the parameter space to better understand the origin of various phases.  Figure~\ref{fig:009d}(a) shows the cuts in $t'-t''$-space associated with reference families $a$ ($t''=0$) [red line] and $b$ ($t''=-t'/2$) [blue line], where the light-green shaded segments show the range over which $\xi$ is significantly depressed, Fig.~M2.  The dotted lines indicate ground state transitions at $x=0$, from $(\pi,\pi)$ to $(\pi,\pi-\delta)$ to $(\delta,\delta)$ order.  The densities-of-states (DOS) shown for the two families in Figs.~\ref{fig:009d}(b) and~\ref{fig:009d}(c) provide insight into the origins of the crossovers.  For reference family $a$ the DOS reveals two crossovers associated with a change from 2D to 1D nesting -- note that the DOS at $t'=-t/2$ shows the characteristic $1/\sqrt(E)$ divergence expected for a 1D system.  Near $t'=-0.5t$, the DOS has two peaks, one at the saddle-point VHS found in 2D materials, the other at the leading edge, as found in 1D systems.  For $t'<-0.46t$, the leading edge is larger, suggestive of weakly-coupled chains, while for $t'>-0.28t$ the leading edge peak turns into a local minimum.  This quasi-1D behavior is suggestive of the nematic phase found in cuprates.  Remarkably, $t'=-0.28t$ approximately coincides with the C-I transition found in Fig.~M2 to lead to a severe $\xi$ depression.  This explains the extended $T$-range of ANN [$(\delta,\delta)$] nesting found in Fig.~\ref{fig:009}(b), which is not characteristic of cuprates.

The origin of the 1D-behavior can be easily understood.  For reference family $a$, the electronic dispersion has the form $E=-2t(c_x+c_y)-4t'c_xc_y\rightarrow -2t$ when $t'=-0.5t$ and either $k_x$ or $k_y$ = 0.  [Here $c_i = cos(k_ia)$, $i=x,y$.]  That is, at the VHS the FS reduces to just the $x$ and $y$ axes.  This explains the strong role of $t''$, which breaks up this perfect nesting, and shifts the doping of optimal ANN nesting away from the VHS-doping.  This causes the FS to split into two pockets, with the VHS corresponding to the topological transition where the pockets split, inset to Fig.~\ref{fig:009d}(b).  The onset of this transition lies close to the doping for Bi2201 (light blue DOS cur). 

\section{Competition between FS and VHS nesting}
\subsection{Origin of FS and VHS nesting}

Here we attempt to separate the two contributions to the susceptibility, from bulk and Fermi surface, and to clarify the origin of the bulk contribution.  To our knowledge there is no standard way to do this, so we have developed simple schemes. 

Consider the expression for the $\omega=0$ bare susceptibility that enters into the Stoner criterion,
\begin{equation}
\chi_0(q)=\sum_k\frac{f(\epsilon_{k+q})-f(\epsilon_k)}{\epsilon_k-\epsilon_{k+q}}.
\label{eq:0a4}
\end{equation}
The denominator is largest if $\epsilon_k=\epsilon_{k+q}$, but in this case the numerator vanishes unless $\epsilon_k$ is at the Fermi level.  Even in this case, the term contributes little weight to the susceptibility integral unless the $q$-shifted FSs are tangent, which is generally satisfied if $q=2k_F$, giving rise to a folded map of the FS -- the ridge of FS nesting.  

This can be seen by comparing a map of $\chi_0'$, Fig.~\ref{fig:01}(a), with the corresponding map of $\chi_0''$, Fig.~\ref{fig:01}(b).  The plot is appropriate for a related cuprate, Bi$_2$Sr$_2$CuO$_6$ (Bi2201), at $x=0.13$ and $\omega = 5$~meV,  
showing in the process that VH nesting also produces a significant effect in the Bi-cuprates.  
Due to the energy $\delta$-function, $\chi_0''$ contains mainly the FS contribution\cite{Mazin}. [The extra feature at $\Gamma$ is the DOS contribution.]   One sees a very similar `inverted-$\gamma$' shaped ridge in both $\chi_0'$ and $\chi_0''$, which is a $q=2k_F$ map of the Fermi surface\cite{Gzm1}.  In Fig.~\ref{fig:01}(d), we plot $\chi_0'$ (red line) and $\chi_0''$ (blue line) along a cut $\Gamma\rightarrow (\pi,0)\rightarrow (\pi,\pi)\rightarrow\Gamma$ in momentum space.  It can be seen that, away from $\Gamma$, the peaks in both components fall at the same $q$-values, and the relative weights of the peaks is comparable.  Thus, the anti-nodal peaks near $\Gamma$ are stronger than the near-$(\pi,\pi)$ peaks.  However, $\chi_0'$ is largest on the near-$(\pi,\pi)$ peaks, since the FS contribution is riding on a large peak associated with the bulk contribution to the susceptibility.

This bulk contribution arises as follows.  Even when $\epsilon_k\ne\epsilon_{k+q}$, there will be a finite contribution to the susceptibility, reduced by the factor $1/(\epsilon_k-\epsilon_{k+q})$, as long as the two energies lie on opposite sides of the Fermi level (for $T=0$).  This contribution will be especially significant near a VHS, where there is a large local DOS (LDOS).  Indeed, this contribution will remain large even as $x$ is doped away from $x_{VHS}$, since while the LDOS on one side of $E_F$ decreases, that on the other side increases, so that the product remains little changed until the VHS peak is quite far from $E_F$. This bulk part provides a smoothly varying background, with hills that can shift the balance of the FS nesting, and in special cases can lead to commensurate nesting away from the FS nesting vector.  Here we try to quantify the dominant role of VHS nesting in creating this background, thereby providing a deeper understanding of the VHS vs FS nesting competition.

\subsection{Separating FS and VHS nesting contributions}

\begin{figure}
\leavevmode  
\epsfig{file=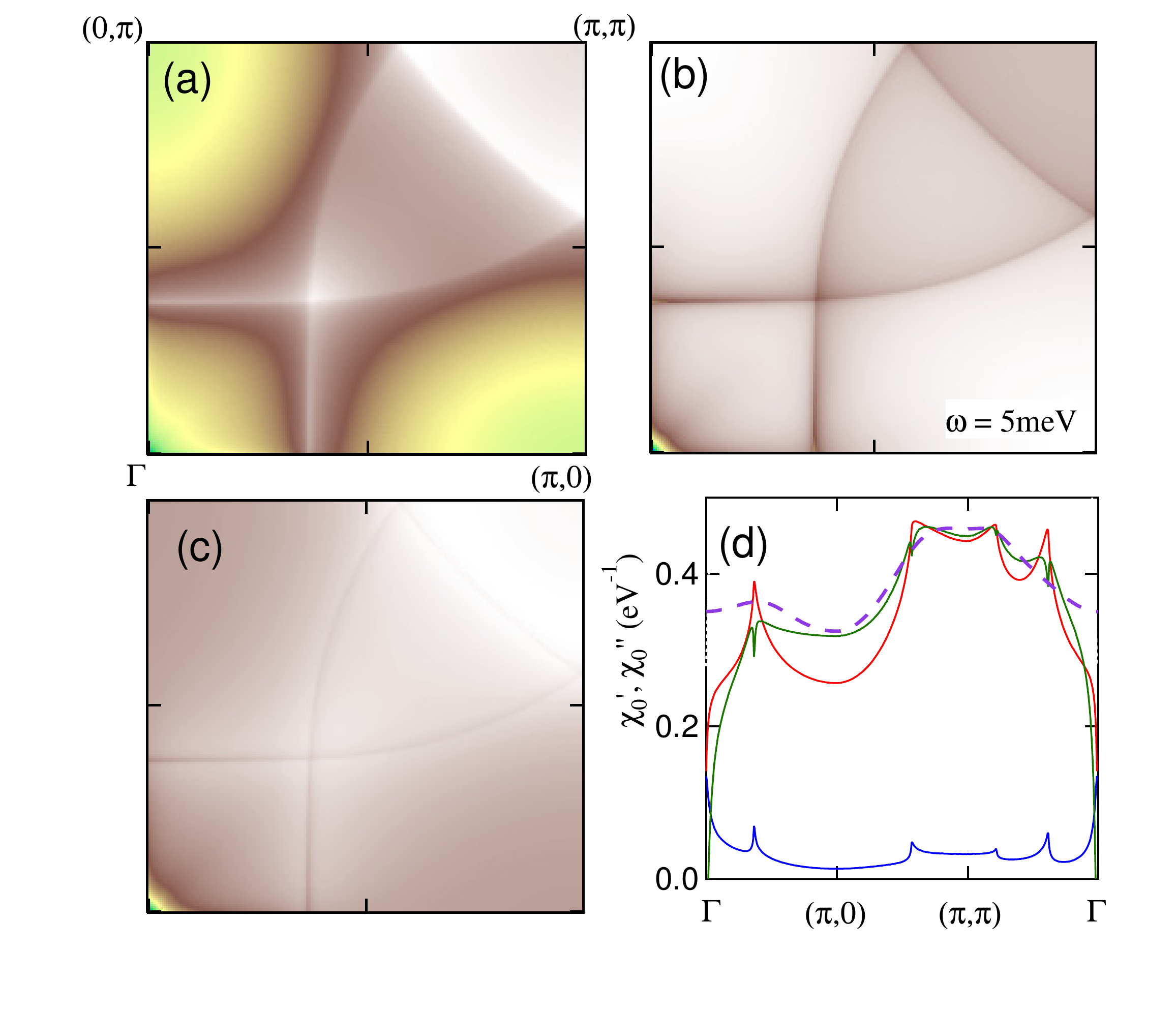,width=9.5cm, angle=0}
\vskip0.5cm
\caption{{\bf Extracting the bulk $\chi_0$ contribution.}
(a) Susceptibility $\chi'_0$ of Bi2201, $x=0.13$, plotted in first Brillouin zone, at $T$ = 0, $\omega=$~5~meV.  Color scale from minimum (blue) to maximum (white). 
(b)  As in (a), but for $\chi''_0$.  (c) Difference plot of (a) and scaled version of (b).
(d) Cuts of susceptibility from earlier frames plotted along high symmetry lines, with red curve representing data in frame (a), blue curve for frame (b), and green curve for frame (c).  Shown also is the curve corresponding to frame a, but at $T=1000$K (violet long-dashed line) scaled by 1.4 and shifted by -0.045~eV$^{-1}$ to match other curves near $(\pi,\pi)$..
} 
\label{fig:01} 
\end{figure}

Two approaches are taken to eliminate the FS nesting features to see the background susceptibility by itself.  First, raising $T$ (violet long-dashed line in Fig.~\ref{fig:01}(d)) washes out the sharp FS features first, while leaving the background features behind.  
Secondly, in frame (c), we plot the difference between $\chi_0'$ and suitably scaled quantity proportional to $\chi_0''$; this is also plotted as the green curve in frame (d).  Since $\chi_0'$ and $\chi_0''$ are Kramers-Kr\"onig transforms, they have different lineshapes, but since the FS map is nearly one-dimensional, a large part of it is still cancelled in taking the difference.  From frame (d) it can be seen that both approaches lead to similar results for the bulk susceptibility, and that this contribution provides the dominant contribution to virtually the whole $(\pi,\pi)$ plateau structure. 
%We show in Section II.B that at high $T$ the dominant fluctuations evolve toward the VHSs of the critical end points of the $t-t'$ model at $t'$ = 0 or -0.5.  

\subsection{Susceptibility Analysis}

\begin{figure}
\leavevmode
\rotatebox{0}{\scalebox{0.38}{\includegraphics{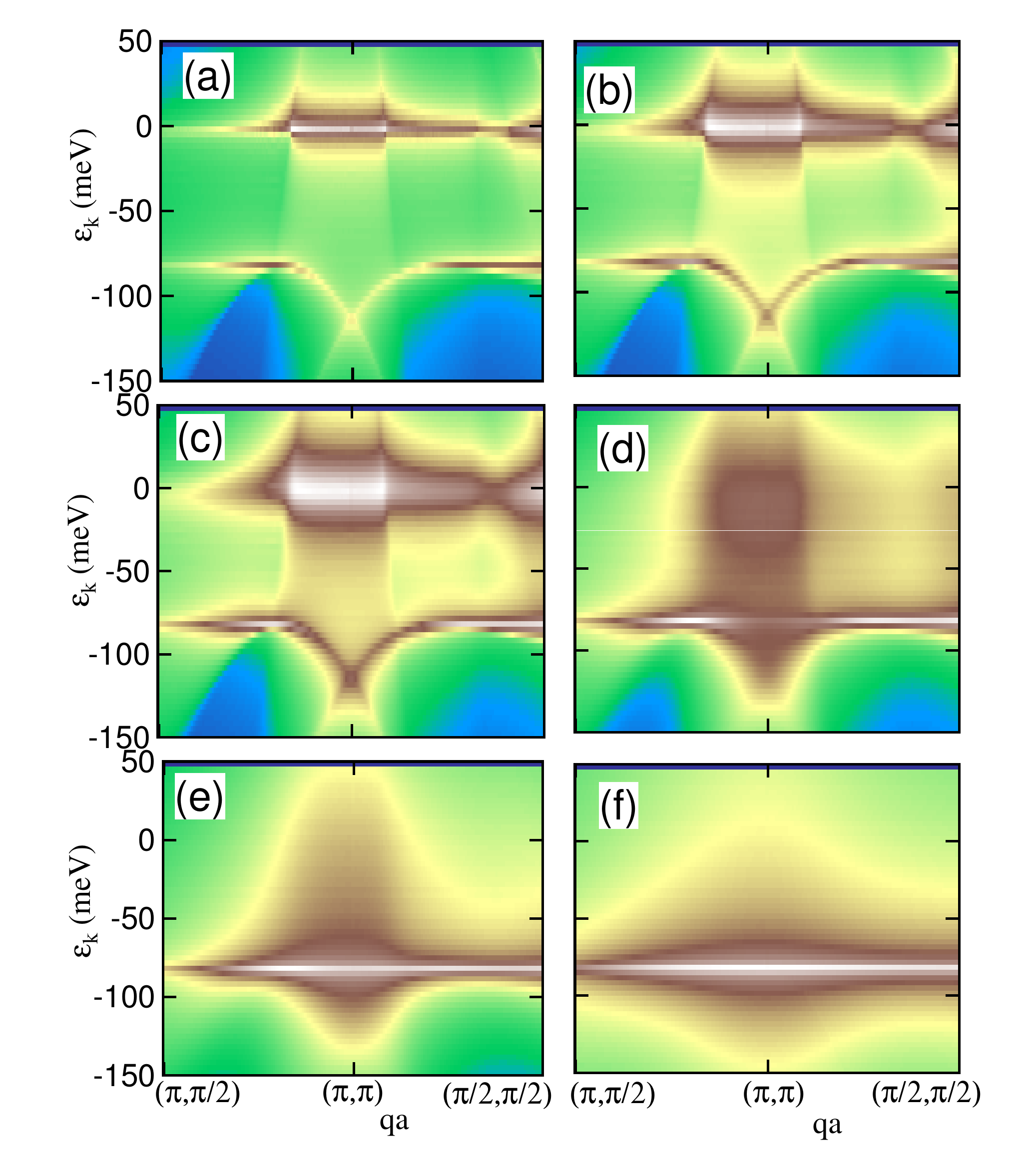}}}
\vskip0.5cm  
\caption{%(Color online) 
{\bf Coherent-incoherent crossover at x=0: role of VHS nesting.}
Deconvolution of the susceptibility for $k$-states in the $(\pi,0)$-patch at several temperatures $T$.  Contribution to the susceptibility of $\chi_{0,2}(q,\epsilon_k)$, Eq.~\ref{eq:0a4b}, is plotted along $q$-cuts near $(\pi,\pi)$ with the color scale ranging from white for a large contribution to blue for negligible contribution for: $T$= 0 (a), 25K (b), 50K (c), 250K (d), 500K (e), and 2000K (f).  
}
\label{fig:2c}
\end{figure}

Finally, we demonstrate the connection between the bulk contribution to $\chi_0$ and the $(\pi,\pi)$ bosonic-VHS (b-VHS), by deconvolving the bare susceptibility, Eq.~\ref{eq:0a4}, into its various $\epsilon_k$ components.  To understand this procedure, recall that the VHS in the electronic dispersion arises when the Fermi surface crosses the Brillouin zone boundary near $(\pi,0)$ and related points, crossing from hole-like to electron-like.  The electron-hole pairs associated with the VHS come in two kinds: intra-VHS pairs, where both carriers lie near the same VHS, so their relative momentum is near $q=(0,0)=\Gamma$, and inter-VHS pairs, where an electron near $(\pi,0)$ and a hole near $(0,\pi)$ have relative momentum near $(\pi,-\pi)$.  The latter pairs carry more weight, forming a b-VHS near $q=(\pi,\pi)$.  Thus, to confirm that the susceptibility peak at $(\pi,\pi)$ is assiciated with the b-VHS, we analyze Eq.~\ref{eq:0a4} for $\chi_0$, assuming that the electron is associated with $\epsilon_k$ and the hole near $\epsilon_{k+q}$.  We must demonstrate that, when $q\sim (\pi,\pi)$, the dominant contribution to $\chi_0$ has $k$ near the VHS at $(\pi,0)$ [or $(0,\pi)$] and $\epsilon_k$ near $E_{VHS}$. 

We divide the Brillouin zone into four patches, around the $\Gamma$, $(\pi,0)$, $(0,\pi)$, and $(\pi,\pi)$ points, and calculate separately the susceptibility due to each patch.  In LSCO, the plateau is predominantly composed of equal contributions from the $(0,\pi)$ and $(\pi,0)$ patches, as expected for a VHS.  We then explore the distribution of values of $\epsilon_k$ which contribute to the susceptibility in the $(\pi,0)$ patch at $x=0$, Fig.~\ref{fig:2c}, rewriting Eq.~\ref{eq:0a4} as
\begin{equation}
\chi_0(q)=\sum_{i=1,4}\sum_{\epsilon_k}\chi_{0,i}(q,\epsilon_k),
\label{eq:0a4b}
\end{equation}
where $i$ runs over the four quadrants $\Gamma$, $(\pi,0)$, $(0,\pi)$, and $(\pi,\pi)$.  

\begin{figure}
\leavevmode  
\epsfig{file=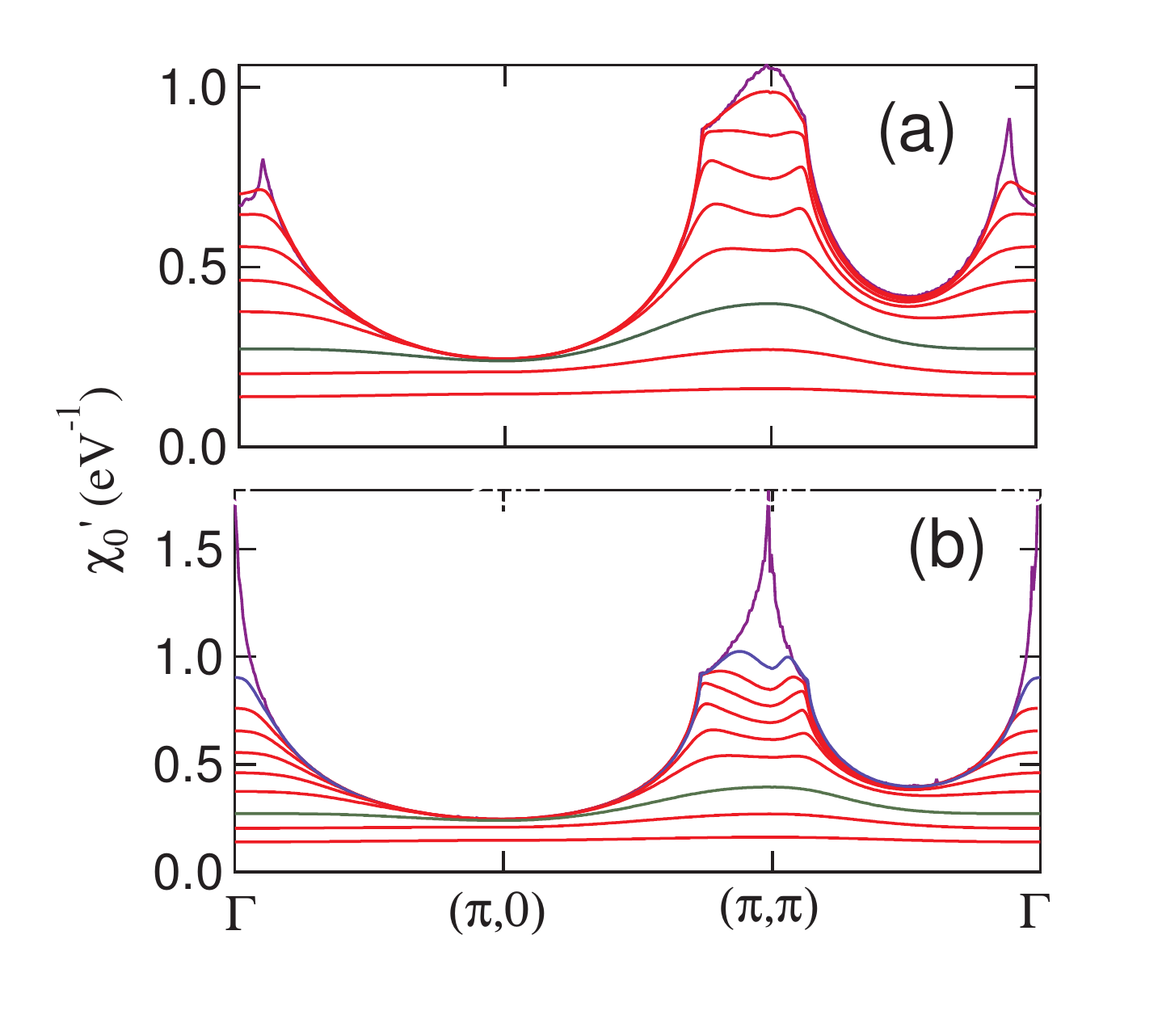,width=8cm, angle=0}
\vskip0.5cm
\caption{
{\bf Lindhard susceptibility of LSCO near the VHS doping.} $\chi_0'(q,\omega=0)$ is plotted for a series of temperatures (from top to bottom) $T$ = 0 (violet), 10 (blue -- frame (b) only), 25, 50, 100, 200, 400, 1000 (green), 2000, and 4000K at doping $x$ = 0.18 (a) and 0.20 = $x_{VHS}$ (b).  
%The $T=1000K$ susceptibility is plotted as a green line. 
} 
\label{fig:0a} 
\end{figure}

There is a clear crossover from a dominant near-FS contribution ($\epsilon_k=0$) at low $T$ to a dominant contribution at high-$T$ coming from the VHS, caused by the breakdown of Pauli blocking, Fig.~M1(b).  At high $T$, the dominant contribution lies along a flat line centered at energy proportional to $k_BT_{VHS}=E_F-E_{VHS}$.  The same crossover is evident in the susceptibility, as the coherent, FS-related features present at low $T$ are washed out, leaving only a featureless (incoherent) $(\pi,\pi)$ plateau at high $T$, Fig.~\ref{fig:0a}.
As doping increases, the same pattern is repeated, but with the flat line moving toward $E_F$ as $T_{VHS}$ decreases, and the crossover from coherent to incoherent shifts to lower $T$, consistent with the pink shaded region in Fig.~M1(c).  The boundaries of this shaded region are $T_{VHS}/2$ and $T_{VHS}/3$, where $T_{VHS}$ is plotted as the dark red dashed curve in  Fig.~M1(c). 

While this explains what happens at $T>T_{VHS}$, the question remains why the broad peak at $(\pi,\pi)$ persists to lower $T$ [e.g., Figs.~\ref{fig:009}(a,c,e) at $x$=0].  To understand this we first look at what happens when Pauli blocking is ostensibly absent -- when $E_F=E_{VHS}$, Fig.~\ref{fig:0a}(b).  While the VHS peak is clearly present at $T=0$ (violet line), finite temperature does not lead to simple thermal broadening but actually to a split peak.  This is a thermally-induced Pauli blocking effect: for $|\epsilon_{\bf k}|$, $|\epsilon_{\bf k+q}|<T$, $f_{\bf k}=f_{\bf k+q}=1/2$ and $\Delta f$ = 0.  Since the $(\pi,\pi)$ susceptibility is pinned at $\omega = 0$, the Pauli blocking turns on first at $(\pi,\pi)$ and then spreads at higher $T$.  This has a number of consequences: (1) As $x$ shifts toward half filling, the peak splitting turns on at higher $T$s, Fig.~\ref{fig:0a}(a) -- this is the lower branch of the $(\pi.\pi)-(\pi,\pi-\delta)$ crossover, plotted in Fig. M1(c).  (2) This is the origin of the ring phase, discussed above in relation to Figs.~\ref{fig:009}(d,f), where the peak susceptibility shifts from $(\pi,\pi)$ to a ring of states encircling $(\pi,\pi)$. (3)  Finally, since the integral of $f$ is fixed by the net doping, when a thermal blocking turns on, some other states will be unblocked at nearby energies.  Thus, the persistence of the net $(\pi,\pi)$ peak at $T<T_{VHS}$ still reflects the underlying presence of the VHS, but is more and more controlled by states away from $(\pi,\pi)$ and closer to $E_F$.  This evolution can be clearly traced in Fig.~\ref{fig:2c}.  It is in this sense that we can refer to this VHS as hidden.

The low-$T$ fadeout of the b-VHS is controlled by the Pauli blocking factor, 
\begin{equation}
\Delta f_{k,Q}=\frac{\sinh{x_-}}{\cosh{x_-}+\cosh{x_+}},
\label{eq:0a5}
\end{equation}
where $x_-=\epsilon_{Q,-}(k)/k_BT$, $x_+=(\epsilon_{Q,+}(k)-E_F)/k_BT$. When the electronic VHS is at the Fermi level, $x_+=0$ and $\Delta f_{k,Q}=\tanh{x_-/2}$, while at lower doping the b-VHS is exponentially suppressed, $\cosh{x_+}\sim exp(T_{VHS}/T)/2$, which equals 1 when $T\sim T_{VHS}/ln(2)$.

\section{Interpretation of Self-consistency}
\subsection{Relation to Bose-Einstein Condensation}

In the conventional picture of Bose-Einstein condensation (BEC), the total number of bosonic modes must satisfy
\begin{equation}
N=\int d\omega n_{BE}(\omega)D_B(\omega),
\label{eq:E1}
\end{equation}
where $n_{BE}$ is the BE distribution function and $D_B$ is the bosonic density of states.  As $T$ decreases, the bosonic chemical potential $\mu$ must adjust to maintain the equality.  However, the largest $\mu$ can be is zero, for which $n_{BE}\rightarrow n_P$, the Planck distribution, and the RHS of Eq.~\ref{eq:E1} takes a fixed value.  For lower $T$ the equation can only be satisfied by putting a macroscopic number of bosons into the lowest energy mode.  

In the present situation, $D_B=\sum_q\chi''$, and $n_P$ can be approximated as $\sim T/\omega$, in which case Eq.~\ref{eq:E1} becomes
\begin{equation}
N\simeq T\sum_q\int d\omega \frac{\chi''(q,\omega)}{\omega} =\pi T\sum_q\chi'(q,\omega=0).
\label{eq:E2}
\end{equation}
Comparing this to Eq.~5 in the main text, self-consistency has produced a number $N_{eff}=N\pi\lambda /A_0$ of excitons which can subsequently Bose condense.  Whereas for 3D BEC the macroscopic occupation of the lowest bosonic mode must be added by hand, in 2D the self-consistent condition automatically causes the occupation of that mode to diverge as $T\rightarrow 0$.

\subsection{Entropy and Dissipation}
As originally noted by McMillan, anomalously large ratios of $2\Delta/k_BT_c$ for any phase transition are generically associated with a dominance of bosonic entropy at finite $T$, associated with low-lying bosonic fluctuations.  But from the fluctuation-dissipation theorem, the net fluctuations at a given frequency $\omega$ are equal to the dissipation $D(\omega)$ at that frequency,
\begin{eqnarray}
D(\omega)= \sum_q \chi''(q,\omega)(n_{BE}(\omega)+1/2)
\nonumber \\
\rightarrow T\sum_q \frac{\chi''(q,\omega)}{\omega}, 
\label{eq:E3}
\end{eqnarray}
so the right-hand side of Eq.~\ref{eq:E2} is just the total dissipation at temperature $T$.  In an ordered phase near $T=0$, the dominant fluctuations will be associated with Goldstone modes, but to show this in a calculation requires extending the present results to full self-consistency between self energy and vertex corrections, as in a parquet calculation.

\section{Mode coupling in Bi2201}
\begin{figure}
\leavevmode
\rotatebox{0}{\scalebox{0.34}{\includegraphics{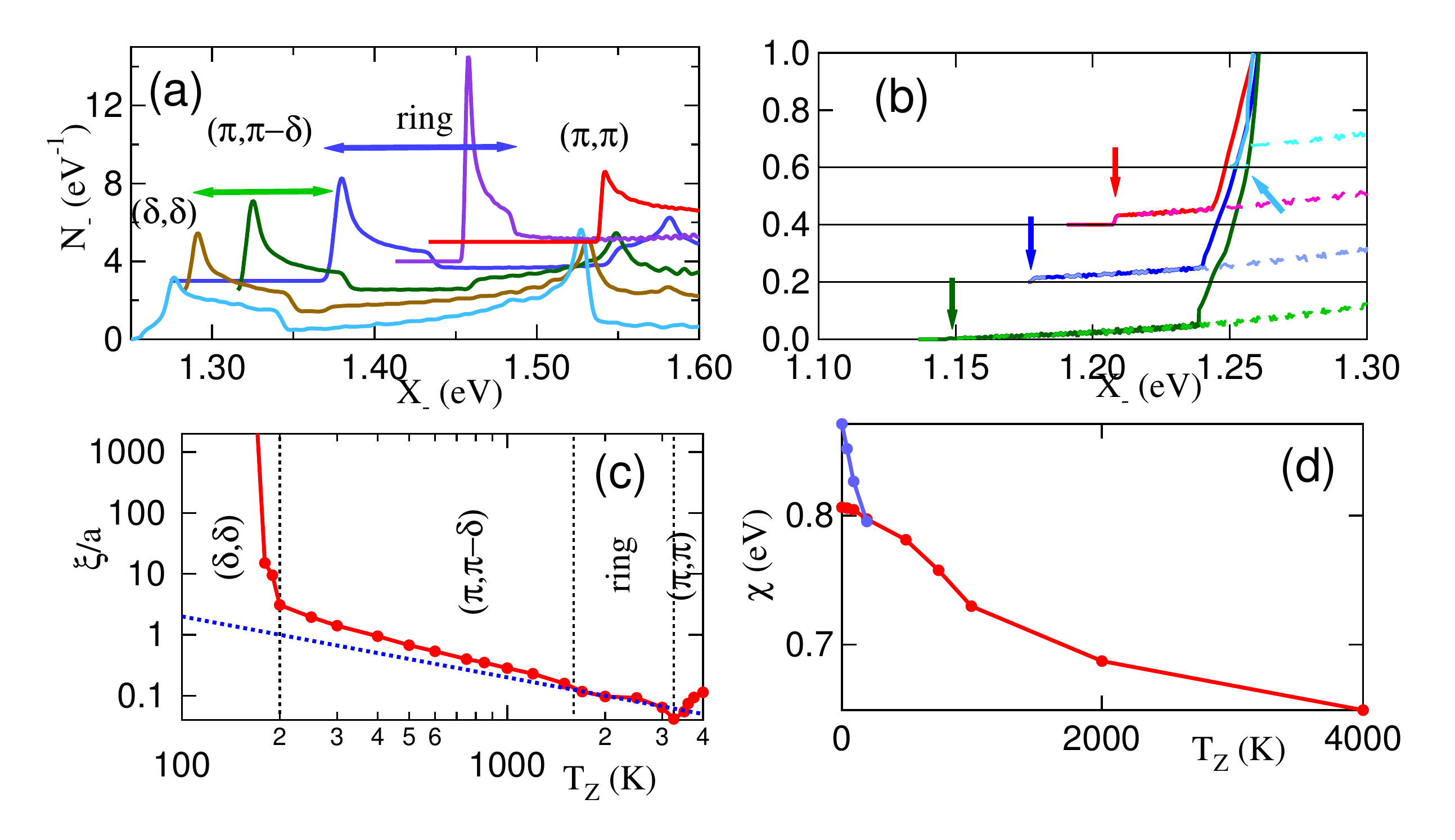}}}
\vskip0.5cm  
\caption{%(Color online)  
{\bf Structures in Bi2201 correlation lengths.}
Temperature dependence of: (a) SDOS $N_{-}$ for $t-t'-t''$ reference state of Bi2201, $x=0.20$. Data sets are offset for clarity; from highest to lowest, $T_Z$ = 4000 (red), 2000 (violet), 1000 (blue), 750 (green), 500 (brown), and 200K (light blue).  (b) Continuation of SDOS in (a) to lower $T_Z$, showing evolution of leading edge.  From top to bottom: $T_Z$ = 200 (light blue), 100 (red), 50 (blue), and 10K (green).  For each curve, a partial SDOS near the ANN peak is superposed to show the SDOS associated with the ANN peak whose onset is marked by arrows.  (c) Correlation length vs $T_Z$.  Blue dotted line represents $\xi\sim T^{-1}$. (d) Susceptibility of $\chi (q)$ peak vs $T_Z$ for $q$ near the $(\pi,\pi)$-plateau (red line) or for ANN nesting (blue line).
} 
\label{fig:3b}
\end{figure}

In Bi2201 and most other cuprates, a larger ratio $|t'/t|$ expands the $(\pi,\pi)$-plateau, leading to a rich phase diagram, Fig.~\ref{fig:009c}(b).  In this case the b-VHS is dominant only at high $T$, while FS-related susceptibility features cause a crossover of the $(\pi,\pi)$ susceptibility from a maximum at high $T$ to a local minimum at lower $T$.  This leads to a different kind of strong mode coupling at intermediate $T$, best exemplified by the ring phase, Figs.~\ref{fig:009}(c,d).   In this phase, the leading-edge SDOS has a `Mexican hat'-like divergence [Main Section II.C], Fig.~\ref{fig:3b}(a) near 2000K, leading to a slow growth in $\xi\sim T^{-1}$, Fig.~\ref{fig:3b}(c). Near the onset of ring order, $T_Z=3250K$, the correlation length has a nonmonotonic $T$-dependence, similar to that seen in LSCO, Fig.~M2(b).

While many pseudogap-like features are similar to LSCO, Bi2201 has some novel properties associated with the ANN instability, which becomes a CDW when charge order is included in the model.  Since the ANN susceptibility peak has a much larger intrinsic curvature, mode coupling effects are weak.  Hence the ANN susceptibility peak grows rapidly at low temperature, and as doping increases, it can surpass the $(\pi,\pi)$-peak, thereby sharply cutting off the growth of near-$(\pi,\pi)$ fluctuations. This is illustrated in Fig.~\ref{fig:3b} for $x$ = 0.20.   Notably, since this peak is not associated with strong mode coupling it has a weak SDOS with a very strong $T$-dependence, Figs.~\ref{fig:3b}(b) and~\ref{fig:3b}(d), crossing the NNN peak near $T_Z=200$K.  The large subsequent growth of the ANN correlation length would ensure the rapid establishment of a true long-range 3D ANN order.  However, corresponding to the large $T$-dependence of $\chi_{ANN}$ there is a similar strong disorder dependence -- particularly in the presence of short-range near-$(\pi,\pi)$ order, which should cause the ANN correlation length to saturate.  Nevertheless, the transition should be manifest as a coherent-incoherent crossover in the overdoped regime, which shifts to higher $T$ as doping increases.  Remarkably, just such a feature has been found in ARPES studies.\cite{Norm1}  Combining this crossover with the $T^*$ crossover at lower doping produces the V-shaped boundary of the `strange metal' phase associated with linear-in-$T$ resistivity.

\section{VHS Excitonic Instability and Kondo-like Physics}

The physics of semimetals is strongly influenced by the simultaneous presence of both holes and electrons, and the attractive interaction between them.  In lightly doped semimetals this can lead to exciton condensation, an extension of the excitonic instability found in small gap semiconductors\cite{ExIns1,ExIns2}.  [See Ref. \onlinecite{graphex} for a recent experimental example.]  While screening modifies excitons in semimetals and doped semiconductors, excitonic effects do not disappear, but shift from $\delta$-function bound excitons to excitonic resonances with a finite width, which persist to high dopings and show strong, even resonant interaction with phonons\cite{Mah2}.  Indeed, Mahan has proposed that exciton formation in a doped semiconductor is resonantly enhanced by the presence of a sharp Fermi surface, in analogy with the edge singularity in x-ray spectra.\cite{Mah1}

The strong effect of excitons in doped semiconductors raises new questions about the role of VHSs in metals.  At a saddle-point VHS in a two-dimensional (2D) metal, or an $M_1$ VHS in a 3D metal\cite{Kittel}, the topology change of the Fermi surface (FS) leads to a large coexisting population of electrons and holes.  Indeed, resonant {\it interband} excitons associated with the $M_1$ VHS (VH excitons) are known to dominate the optical spectra of metals.  Similarly, in cuprates, an {\it intraband} Van Hove exciton can arise near an electronic saddle-point VHS\cite{ExIns3,ExIns4}.  Just as in conventional exciton Bose condensation, the mean-field transition corresponds to the $T$ where [quasi]excitons form, but since these are nearly localized, there is a large spread in $q$, leading to incoherent VHS nesting.  The real transition lies at much lower $T$, when all pairs condense into the lowest energy $q$-state.
This excitonic VHS continues to play a role even when the electronic VHS is away from $E_F$, leading to a peak in $\chi_0'$ when either $T$ or $\omega$ is $\sim E_F-E_{VHS}$.  This is due to scattering between states near the VHS and states near $E_F$, and is similar to Kondo physics, with the VHS states playing the role of $f$-electrons.  

To clarify the connection to Kondo physics, we write a variational ground state for the singlet exciton as
\begin{equation}
|\psi_G>=\prod_{k,\sigma}\Bigl(u_k+v_kc^{\dagger}_{k+Q,\sigma}c_{k,\sigma}\Bigr)|F_{(\pi,\pi)}>,
\label{eq:B1}
\end{equation}
where $c^{\dagger}_{k,\sigma}$ [$c_{k,\sigma}$] are creation [annihilation] operators for an electronic state of momentum $k$ and spin $\sigma$, and $u_k$, $v_k$ are variational parameters.  Here 
%$\bar\sigma =-\sigma$ and 
$|F_{(\pi,\pi)}>$ is a filled Fermi sea of the first $(\pi,\pi)$-superlattice Brillouin zone, and the excitonic order parameter is $\Delta_{CDW}=<c^{\dagger}_{k+Q,\uparrow}c_{k,\uparrow}+c^{\dagger}_{k+Q,\downarrow}c_{k,\downarrow}+c.c.>$.  This model is a direct generalization of Yosida's variational calculation of the screening of a Kondo impurity to a single-band Kondo lattice model.\cite{Yosi,Mah}  Moreover, to form a localized exciton wave-packet, the restriction $q=Q$ in Eq.~\ref{eq:B1} must be dropped, and many $q$-vectors summed over, accounting for the bosonic entropy of the model -- indeed the SDOS quantifies the number of near-$Q$ states that contribute to the entropy.  
Since the Fermi velocity $v_F=0$ at a VHS, these excitons are nearly localized.  
%Moreover, since the lifetime of an electron with well-defined spin can be greatly extended by binding it to an opposite-spin hole, the VHS singlet exciton provides a model for local moment formation. Hence, we have a picture of a transition from a metal to a Mott insulator.  
This reinforces the connection of the coherent-incoherent crossover to Kondo physics\cite{CFP}: in Main Fig.~M2(b), region I corresponds to the buildup of a Kondo lattice of nearly localized spins, while region II is a regime of crossover to itinerant physics.

\section{Relation to Other Calculations}

A comparison between our QPGW model and other `DFT+' calculations was given in Ref.~\onlinecite{AIP}; here we discuss this comparison specifically in regard to pseudogap physics.
A number of groups are working on MBPT-type extensions of DFT calculations.  However, DMFT and its cluster extensions are limited by a very low resolution in momentum $q$.  In particular, this means that even if they capture Mermin-Wagner physics, they are typically limited to $\xi<7a$\cite{MaJar} [Fig. 50 of Ref.~\onlinecite{AIP}].  Furthermore, in averaging over large patches in momentum space, they lose the ability to resolve ordinary nesting instabilities, such as the ANN instability responsible for cuprate CDWs [Fig. 49 of Ref.\onlinecite{AIP}].  On the other hand, the fact that they can see the pseudogap, exen when averaging over a quarter of the Brillouin zone is consistent with our finding that the pseudogap is spread over many separate modes in $q$.

However, some DMFT extensions [dynamic vertex approximation\cite{DGA1,DGA3,DGA2}, dual fermion\cite{DF1,DF2}, and one-particle irreducible approaches\cite{OPI}] seek to calculate the two-particle vertex, which allows them to calculate the momentum-dependent self energy, and these approaches are able to reproduce Mermin-Wagner physics and vertex corrections.  
We note in particular that a recent dynamical cluster approximation (DCA)\cite{flucdi} calculation sorted out the contribution of various channels of two-particle scattering to the self energy (`fluctuation diagnostics'), and found that spin fluctuations are the origin of the pseudogap, while charge and pairing fluctuations play a marginal role, consistent with our results.  This has also been found experimentally\cite{Badoux}.

We further note that the parquet equations\cite{parq1,parq2} could be quite useful, if the recently discovered divergences\cite{flucdi} can be overcome. The mutual self-consistency between the self-energy and vertex corrections will be necessary to extend the current results to low temperatures.

Finally, in many strong-coupling calculations, it is implicitly assumed that Fermi surface nesting is irrelevant\cite{RVB}, and there is a special role for commensurate $(\pi,\pi)$ magnetic order.  Thus, in the Yang-Rice-Zhang phenomenological model\cite{YRZ}, the nesting is tied to the magnetic Brillouin zone boundary.  The commensurate phase we have found provides a clear DFT-based explanation for that effect, as well as a clear picture of how this evolved into a more Fermi-liquid-like phase at higher doping or larger $t'$.

After our paper was posted [arXiv:1505.04770], we became aware of another paper\cite{Pepin} that also finds that strong bosonic fluctuations over a `quasidegenerate distribution of $2{\bf p_F}$ ordering wave vectors' drives the pseudogap, although they find it is predominantly in the charge sector, forming a `resonant Peierls excitonic state'.